\documentclass[12pt]{iopart}
\usepackage{iopams}
\usepackage{setstack}
\usepackage{amssymb}
\usepackage{latexsym}
\usepackage{color}
\usepackage{graphicx}   
     
\begin{document}
\title[Finite Gaussians and finite quantum oscillators]{Gaussian type functions defined \\ on a finite set and  finite oscillators}
\author{Nicolae Cotfas}
\address{University of Bucharest,  Physics Department,\\ P.O. Box MG-11, 077125 Bucharest, Romania}

\eads{\mailto{ncotfas@yahoo.com}}

\begin{abstract}
The mathematical description of the quantum harmonic oscillator is essentially based on the Gaussian function. In the case of a quantum oscillator with finite-dimensional Hilbert space, the position space consists in a finite number of points, and the ground state can be regarded as a finite version of the Gaussian function. Conversely, each finite Gaussian can be used in order to define some finite oscillators. We investigate certain finite Gaussians as concern their Fourier transform, Wigner function and associated oscillators.
\end{abstract}

\section{Introduction}
In quantum mechanics and quantum information there exists an increasing interest in quantum systems with 
finite-dimensional Hilbert space. The finite-dimensional versions of the mathematical objects used in the continuous case are expected to play an
important role. 
In section 2, we review some fundamental mathematical tools available in a finite-dimensional Hilbert space for the description of quantum systems.
Particularly, we present some results concerning the Kravchuk functions in a version suitable for our purpose.

In the case of a quantum system with finite-dimensional Hilbert space there are several candidates for the role of Gaussian function. In section 3 we investigate five of them as concern their finite Fourier transform and the corresponding Wigner function.
In section 4, each of these finite Gaussians is used in order to generate a finite tight frame representing a finite  counterpart of the canonical system of coherent states. 
In the first part of section 5, we review some known one-dimensional finite oscillators. More examples can be found in \cite{APW}. In the second part of section 5, based on the considered finite Gaussians and the finite frame quantization, we define some new finite oscillators. In section 6, we investigate the occurrence of revivals. Some explicit Gaussians and oscillators are presented in section 7.
 
Our aim is to bring a contribution to the mathematical formalism available for the description of the quantum systems with finite-dimensional Hilbert space.

\section{Mathematical tools available in a finite dimensional Hilbert space}

Let $j\!\in \! \{ 1,2,3,...\}$ and $d\!=\!2j\!+\!1$. Any $d$-dimensional Hilbert space is isomorphic to $\mathbb{C}^d$, and can be regarded as the space of all the complex functions defined on a set with $d$ elements. We consider 
\[
\mathbb{C}^d=\{ \ \psi\!:\!\{-j, -j\!+\!1,\, ...\, ,j\!-\!1, j\}\longrightarrow \mathbb{C}\ \ |\ \ \psi \ \mbox{is a function}\ \}
\]
with the scalar product defined as
\begin{equation}\label{scalprod}
\langle \varphi ,\psi \rangle =\sum_{n=-j}^j \overline{\varphi(n)}\, \psi(n).
\end{equation}
In order to define certain mathematical objects, we have to embed $\mathbb{C}^d$ in the space of all the functions of the form $\psi:\mathbb{Z}\longrightarrow \mathbb{C}$ by identifying it either with the space 
\begin{equation}
\ell ^2(\mathbb{Z}_d)\!=\!\{ \ \psi \!:\!\mathbb{Z}\!\longrightarrow \!\mathbb{C}\ \ |\ \ \psi(n\!+\!d)\!=\!\psi(n)\ \ \mbox{for any}\ \ n\!\in \!\mathbb{Z}\ \}
\end{equation}
of periodic functions of period $d$ or with the space 
\begin{equation}
\ell ^2[-j,j]\!=\!\{ \, \psi \!:\!\mathbb{Z}\!\longrightarrow \!\mathbb{C}\ \, |\ \, \psi(n)\!=\!0\ \ \mbox{for}\ \ n\!\not\in \!\{-j, -j\!+\!1,\, ...\, ,j\!-\!1, j\}\, \}
\end{equation}
of all the functions null outside $\{-j, -j\!+\!1,\, ...\, ,j\!-\!1, j\}$.
Here $\mathbb{Z}_d$ is the ring of integers modulo $d$.
Depending on the used identification, a function $\psi\!\in\!\mathbb{C}^d$ is extended up to a periodic function with period $d$ or up to a function null outside $\{-j, -j\!+\!1,\, ...\, ,j\!-\!1, j\}$.
Conversely, any function $\psi$ from $\ell ^2(\mathbb{Z}_d)$ or 
$\ell ^2[-j,j]$ coresponds uniquely to a function $\psi\!\in\!\mathbb{C}^d$, which can be obtained by taking the restriction to 
$\{-j, -j\!+\!1,\, ...\, ,j\!-\!1, j\}$. The scalar product in all the three spaces $\mathbb{C}^d$, $\ell ^2(\mathbb{Z}_d)$ and $\ell ^2[-j,j]$ is defined by the same formula, namely (\ref{scalprod}). The Hilbert spaces $\mathbb{C}^d$, $\ell ^2(\mathbb{Z}_d)$ and $\ell ^2[-j,j]$ are isomorphic, but they offer distinct formal advantages.

The evolution of a quantum particle along an axis is described by the Hilbert space
\begin{equation}
L^2(\mathbb{R})=\left\{ \ \psi :\mathbb{R}\longrightarrow \mathbb{C}\ \left|
\ \int_{-\infty }^\infty |\psi(q)|^2\, dq <\infty \right. \right\}
\end{equation}
with the scalar product defined as
\begin{equation}
\langle \varphi ,\psi \rangle =\int_{-\infty}^\infty \overline{\varphi(q)}\, \psi(q)\, dq.
\end{equation}
Similarly, the evolution of a quantum particle along $\{-j, -j\!+\!1,\, ...\, ,j\!-\!1, j\}$ is usually described by using one of the Hilbert spaces $\mathbb{C}^d$, $\ell ^2(\mathbb{Z}_d)$ and $\ell ^2[-j,j]$.\\
The {\em Fourier transform} $F\!:\!\mathbb{C}^d\!\longrightarrow \! \mathbb{C}^d:\psi \!\mapsto \!F[\psi]$, where
\begin{equation}\label{defF}
F[\psi ](k)=\frac{1}{\sqrt{d}}\sum_{n=-j}^j {\rm e}^{-\frac{2\pi {\rm i}}{d}kn}\, \psi(n),
\end{equation}
is a unitary transformation. Its inverse is the adjoint transformation
\begin{equation}
F^+[\psi ](k)=\frac{1}{\sqrt{d}}\sum_{n=-j}^j {\rm e}^{\frac{2\pi {\rm i}}{d}kn}\, \psi(n).
\end{equation}
We have $F^2[\psi](k)\!=\!\psi(-k)$, and $F^+[\psi]\!=\!F[\psi]$ for any even function $\psi $. The Fourier transform $F\!:\!\ell ^2(\mathbb{Z}_d)\!\longrightarrow\!\ell ^2(\mathbb{Z}_d)$ corresponding to $F\!:\!\mathbb{C}^d\!\longrightarrow \! \mathbb{C}^d$ is also defined by (\ref{defF}).

The functions $\{ \delta_k\}_{k=-j}^j$, defined by the relation 
\begin{equation}\label{defbasis}
\delta_k (n)\!=\!\delta_{kn}\!=\!\left\{ 
\begin{array}{ccc}
1 & \mbox{for} & n\!=\!k,\\[1mm]
0 & \mbox{for} & n\!\neq \!k,
\end{array} \right.
\end{equation}
form an orthonormal basis in $\mathbb{C}^d$ and $\ell ^2[-j,j]$, called the canonical or computational basis. The corresponding basis in $\ell ^2(\mathbb{Z}_d)$, written as $\{ \delta_k\}_{k\in\mathbb{Z}_d}$, is  defined by the relation
\begin{equation}
\delta_k (n)\!=\!\left\{ 
\begin{array}{ccc}
1 & \mbox{for} & n\!=\!k\ ({\rm modulo}\, d),\\[1mm]
0 & \mbox{for} & n\!\neq \!k\ ({\rm modulo}\, d).
\end{array} \right.
\end{equation}
By writing $|j;k\rangle $ instead of $\delta_k$,    we have $\langle j;k|j;\ell \rangle \!=\!\delta_{k\ell }$ and the resolution 
 \begin{equation}
  \begin{array}{l}
 \sum\limits_{k=-j}^j|j;k\rangle \langle j;k|\!=\!\mathbb{I},
 \end{array}
  \end{equation}
of the identity operator $\mathbb{I}|\psi \rangle \!=\!|\psi \rangle $. 
The functions $\{ F^+[\delta_k]\}_{k=-j}^j$ form a complementary orthonormal basis. By writing $|j;k\rangle\!\rangle $ instead of $F^+[\delta_k]$,    we have
\begin{equation}
\begin{array}{l}
|j;k\rangle\!\rangle \!=\!\frac{1}{\sqrt{d}}\sum\limits_{n=-j}^j\!\! {\rm e}^{\frac{2\pi {\rm i}}{d}kn}\,|j;n\rangle 
\end{array}
\end{equation}
and
 \begin{equation}
 \begin{array}{l}
 \langle\!\langle j;k|j;\ell \rangle\!\rangle \!=\!\delta_{k\ell },
 \end{array}
 \qquad \qquad 
 \begin{array}{l}
 \sum\limits_{k=-j}^j|j;k\rangle\!\rangle\langle\! \langle j;k|\!=\!\mathbb{I}.
 \end{array}
  \end{equation} 
As concern the Fourier transform, we have
\[
\begin{array}{ll}
F\!=\!\frac{1}{\sqrt{d}}\!\!\sum\limits_{n,m=-j}^j\!\! {\rm e}^{-\frac{2\pi {\rm i}}{d}nm}\,|j;n\rangle \langle j;m|,\qquad & 
F\!=\!\frac{1}{\sqrt{d}}\!\!\sum\limits_{n,m=-j}^j\!\! {\rm e}^{-\frac{2\pi {\rm i}}{d}nm}\,|j;n\rangle\!\rangle\langle\! \langle j;m|,\\[2mm]
F^2\!=\!\!\sum\limits_{n=-j}^j\!\! |j;-n\rangle \langle j;n|\!=\!(F^+)^2,\qquad &  
F^2\!=\!\!\sum\limits_{n=-j}^j\!\! |j;-n\rangle\!\rangle\langle\! \langle j;n|\!=\!(F^+)^2.
\end{array}
\]

The self-adjoint operators $Q,P\!:\!\mathbb{C}^d\!\longrightarrow \! \mathbb{C}^d$, where
\begin{equation}
(Q\psi)(n)=n\, \psi(n)\qquad \mbox{and} \qquad P=F^+QF
\end{equation}
admit the spectral decompositions
\begin{equation}
\begin{array}{l}
Q=\sum\limits_{n=-j}^jn\, |j;n\rangle \langle j;n|,\qquad P=\sum\limits_{k=-j}^jk\, |j;k\rangle\!\rangle \langle\! \langle j;k|.
\end{array}
\end{equation}
The Schwinger's unitary operators \cite{Schwinger}
\begin{equation}
A,B\!:\!\mathbb{C}^d\!\longrightarrow \! \mathbb{C}^d, \qquad 
\begin{array}{l}
A={\rm e}^{-\frac{2\pi {\rm i}}{d}P}=\sum\limits_{k=-j}^j{\rm e}^{-\frac{2\pi {\rm i}}{d}k}|j;k\rangle\!\rangle \langle\! \langle j;k|,\\[3mm]
B={\rm e}^{\frac{2\pi {\rm i}}{d}Q}=\sum\limits_{n=-j}^j{\rm e}^{\frac{2\pi {\rm i}}{d}n}|j;n\rangle \langle j;n|
\end{array}
\end{equation}
can be simpler analyzed by using $\ell^2(\mathbb{Z}_d)$ instead of $\mathbb{C}^d$. 
They satisfy the relations \cite{Schwinger,Vourdas}
\begin{equation} 
\begin{array}{ll}
A^\alpha |j;n\rangle=|j;n\!+\!\alpha \rangle, & B^\beta |j;n\rangle={\rm e}^{\frac{2\pi {\rm i}}{d}n\beta}|j;n\rangle ,\\[2mm]
A^\alpha|j;k\rangle\!\rangle ={\rm e}^{-\frac{2\pi {\rm i}}{d}k\alpha}|j;k\rangle\!\rangle ,\qquad  & B^\beta |j;k\rangle\!\rangle =|j;k\!+\!\beta\rangle\!\rangle ,\\[2mm]
 (A^\alpha \psi )(n)=\psi (n\!-\!\alpha ), &    (B^\beta \psi )(n)={\rm e}^{\frac{2\pi {\rm i}}{d}n\beta}\psi (n) ,\\[2mm]           
A^d=B^d=\mathbb{I}, & A^\alpha B^\beta ={\rm e}^{-\frac{2\pi {\rm i}}{d}\alpha \beta}                              
B^\beta A^\alpha  ,                                            
\end{array}
\end{equation}                                                           
for any $\alpha ,\beta \!\in \!\mathbb{Z}$ and $\psi \!\in \!\ell^2(\mathbb{Z}_d)$. The $d^2$ operators
\begin{equation}
D(\alpha ,\beta)\!=\!{\rm e}^{\frac{\pi {\rm i}}{d}\alpha \beta}A^\alpha B^\beta\!=\!{\rm e}^{-\frac{\pi {\rm i}}{d}\alpha \beta}B^\beta A^\alpha ,
\end{equation}
satisfying the relation
\begin{equation}
D(\alpha_1 ,\beta_1)\, D(\alpha_2 ,\beta_2)\!=\!{\rm e}^{-\frac{\pi {\rm i}}{d}(\alpha_1 \beta_2-\alpha_2\beta_1)}D(\alpha_1\!+\!\alpha_2 ,\beta_1\!+\!\beta_2),
\end{equation}
define a projective representation of a finite version of the                                                                                      
Heisenberg-Weyl group.

The convolution of two functions $\varphi ,\psi \!\in \!\ell^2(\mathbb{Z}_d)$ is the function $\varphi \!\ast\! \psi :\mathbb{Z}_d\longrightarrow \mathbb{C}$,
\begin{equation}
(\varphi  \!\ast\! \psi)(n)=\sum_{m=-j}^j\varphi(m)\, \psi(n\!-\!m)=\sum_{m=-j}^j\varphi(n\!-\!m)\, \psi(m)
\end{equation}
and we have
\[
\begin{array}{rl}
F[\varphi  \!\ast\! \psi](k) & \!\!\!\!=\frac{1}{\sqrt{d}}\sum\limits_{n,m=-j}^j{\rm e}^{-\frac{2 \pi {\rm i}}{d}kn}\, \varphi(m)\, \psi(n-m)\\[2mm]
& \!\!\!\!=\frac{1}{\sqrt{d}}\sum\limits_{m,\ell=-j}^j{\rm e}^{-\frac{2 \pi {\rm i}}{d}k(m+\ell)}\, \varphi(m)\, \psi(\ell),
\end{array}
\]
that is,
\begin{equation}
F[\varphi \!\ast\! \psi]=\sqrt{d}\, F[\varphi]\,  F[\psi].
\end{equation}
The {\em discrete Wigner function}  $W_{\!\psi} \!:\!\mathbb{Z}_d\!\times\!\mathbb{Z}_d\!\longrightarrow \! \mathbb{R}$,
\begin{equation}\label{defWigner}
W_{\!\psi} (n,m)\!=\!\frac{1}{d}\sum\limits_{k=-j}^j\mathrm{e}^{\frac{4\pi \mathrm{i}}{d}mk}\, \psi(n\!-\!k)\, \overline{\psi(n\!+\!k)}
\end{equation}
corresponding to a function $\psi \!\in \!\ell ^2(\mathbb{Z}_d)$ satisfies the relation
\[
W_{\!\psi} (n,m)\!=\!\frac{1}{d}\sum\limits_{k=-j}^j\mathrm{e}^{-\frac{4\pi \mathrm{i}}{d}nk}\, F[\psi](m\!-\!k)\, \overline{F[\psi](m\!+\!k)}
\]
which implies that
\[
W_{\!F[\psi]} (n,m)\!=\!\frac{1}{d}\sum\limits_{k=-j}^j\mathrm{e}^{-\frac{4\pi \mathrm{i}}{d}nk}\, \psi(-m\!+\!k)\, \overline{\psi(-m\!-\!k)}.
\]
Particularly, in the case of an even function $\psi $ we have
\begin{equation}\label{FWigner}
W_{\!F[\psi]} (n,m)=W_{\!\psi} (m,-n).
\end{equation}

By admitting for gamma function that
\begin{equation}
\frac{1}{\Gamma (n)}=0\qquad \mbox{for}\ \ n\!\in \!\{ 0,-1,-2,-3, ...\},
\end{equation}
we get an extended definition for the binomial coefficients
\begin{equation}
C_k^n\!=\!\frac{\Gamma (k\!+\!1)}{\Gamma (n\!+\!1)\, \Gamma (k\!-\!n\!+\!1)}\!=\!
\left\{ 
\begin{array}{cll}
\frac{k!}{n!\, (k\!-\!n)!} & \mbox{for} & n\!\in \!\{ 0,1,2, ...,k\},\\[2mm]
0 & \mbox{for} & n\!\in \!\mathbb{Z}\backslash\{ 0,1,2, ...,k\}.
\end{array}\right.
\end{equation}
Particularly, the function
\[
\varphi:\mathbb{Z}\longrightarrow \mathbb{R},\qquad \varphi(n)=C_{2j}^{j+n}
\]
belongs to the space $\ell^2[-j,j]$.
For any $n,m\!\in \!\{-j,-j\!+\!1,...,j\!-\!1,j\}$, the coefficient of $X^{j+m}$ in 
$(1\!-\!X)^{j+n}(1\!+\!X)^{j-n}$ is the polynomial of degree $j\!+\!m$
\[
K_m(n)=\sum_{k=0}^{j+m}(-1)^k\, C_{j+n}^k\, C_{j-n}^{j+m-k}
=\sum_{k=0}^{j+m}(-1)^{j+m-k}\, C_{j-n}^k\, C_{j+n}^{j+m-k},
\]
called {\em Kravchuk polynomial}. We have $K_m(-n)\!=\!(-1)^{j+m}\, K_m(n)$, and 
\begin{equation}\label{defKpoly}
\sum_{m=-j}^{j}K_m(n)\, X^{j+m}=(1\!-\!X)^{j+n}(1\!+\!X)^{j-n}.
\end{equation}
The first three polynomials are: $K_{-j}(n)\!=\!1$, \ $K_{-j+1}(n)\!=\!-2n$, \ 
$K_{-j+2}(n)\!=\!2n^2\!-\!j$.\\[5mm]
{\bf Theorem 1}. {\it The polynomial functions $K_m\!:\!\{-j,-s\!+\!1,...,j\!-\!1,j\}\longrightarrow \mathbb{R}$,
\begin{equation}
\qquad \qquad 
K_m(n)\!=\!\sum_{k=0}^{j+m}(-1)^k\, C_{j+n}^k\, C_{j-n}^{j+m-k}\\[-2mm]
\end{equation}
\mbox{}\qquad \qquad \qquad   satisfy the relation\\[-4mm]
\begin{equation}\label{Krort}\qquad \qquad 
\begin{array}{l}
\frac{1}{2^{2j}}\sum\limits_{n=-j}^j C_{2j}^{j+n}\, K_m(n)\, K_\ell (n)= C_{2j}^{j+m}\, \delta_{m\ell},
\end{array}
\end{equation}
\mbox{}\qquad \qquad \qquad for any  $m,\ell \!\in \!\{-j,-j\!+\!1,...,j\!-\!1,j\}$.}\\[3mm]
{\it Proof}. Direct consequence of the polynomial relation
\[
\begin{array}{l}
\sum\limits_{m,\ell=-j}^j\left(\frac{1}{2^{2j}}\sum\limits_{n=-j}^j C_{2j}^{j+n}\, K_m(n)\, K_\ell (n)\right)X^{j+m}\, Y^{j+\ell}\\[2mm]
\mbox{}\qquad =\frac{1}{2^{2j}}\sum\limits_{n=-j}^j C_{2j}^{j+n}\, \sum\limits_{m=-j}^{j}K_k(n)\, X^{j+m}\, \sum\limits_{\ell=-j}^{j}K_\ell (n)\, X^{j+\ell }\\[2mm]
\mbox{}\qquad  =\frac{1}{2^{2j}}\sum\limits_{n=-j}^j C_{2j}^{j+n}\,(1\!-\!X)^{j+n}(1\!+\!X)^{j-n}(1\!-\!Y)^{j+n}(1\!+\!Y)^{j-n} \\[2mm]
\mbox{}\qquad =\frac{1}{2^{2j}}[(1\!-\!X)(1\!-\!Y)+(1\!+\!X)(1\!+\!Y)]^{2j}= (1+XY)^{2j}
\\[2mm]
\mbox{}\qquad =\sum\limits_{m=-j}^j C_{2j}^{j+m}X^{j+m} Y^{j+m}.\qquad \opensquare
\end{array}
\]
The {\em Kravchuk functions} $\mathfrak{K}_m\!:\!\{-j,-j\!+\!1,...,j\!-\!1,j\}\longrightarrow \mathbb{R}$,
\begin{equation}
\mathfrak{K}_m(n)=\frac{1}{2^j}\sqrt{\frac{C_{2j}^{j+n}}{C_{2j}^{j+m}}}\,K_m(n)
\end{equation}
satisfying the relation $\mathfrak{K}_m(-n)\!=\!(-1)^{j+m}\mathfrak{K}_m(n)$, form an orthonormal basis in $\ell^2(\mathbb{Z}_d)$:
\begin{equation}
 \langle \mathfrak{K}_m|\mathfrak{K}_\ell \rangle =\delta_{m\ell },\qquad \sum_{m=-j}^j|\mathfrak{K}_m\rangle \langle \mathfrak{K}_m|=\mathbb{I}.
  \end{equation}
{\bf Theorem 2}. {\it The {\em Kravchuk functions} $\mathfrak{K}_m$ satisfy the relation
\begin{equation}
\mathfrak{K}_m(n)=\frac{1}{2^j}\sqrt{C_{2j}^{j+m}\, C_{2j}^{j+n}}\,\, {}_2F_1\!\!\left(\left.\!\!\!\!
\begin{array}{c}
-j\!-\!m,\, -j\!-\!n\\
-2j
\end{array}\!\!\right|\!2\right),
\end{equation}
\mbox{}\qquad \qquad \quad and particularly, we have}
\begin{equation}\label{Krsimrel}
\mathfrak{K}_m(n)=\mathfrak{K}_n(m),\qquad \mbox{for any}\ \ n,m\!\in \!\{ -j,-j\!+\!1,...,j\!-\!1,j\}.
\end{equation}
{\it Proof}. It is known that the hypergeometric function
\begin{equation}\label{hipge}
{}_2F_1\left(\left.
\begin{array}{c}
a,\, b\\
c
\end{array}\right|z\right)=\sum_{k=0}^\infty \frac{(a)_k\, (b)_k}{(c)_k}\frac{z^k}{k!}, 
\end{equation}
where $(\alpha)_k\!=\!\alpha(\alpha\!+\!1)...(\alpha\!+\!k\!-\!1)\!=\!\Gamma(\alpha\!+\!k)/\Gamma(\alpha)$, satisfies the relation \cite{Nikiforov}
\[
{}_2F_1\left(\left.\!\!
\begin{array}{c}
-n,\, \beta\\
\gamma
\end{array}\right|z\right)=\frac{\Gamma (\gamma )\, \Gamma(\gamma \!-\!\beta \!+\!n)}{\Gamma (\gamma \!+\!n)\, \Gamma(\gamma\!-\!\beta)}\,
{}_2F_1\left(\left.\!\!
\begin{array}{c}
-n,\, \beta\\
\beta\!-\!\gamma\!-\!n\!+\!1
\end{array}\right|1\!-\!z\right).
\]
Since $(-\alpha )_k\!=\!(-1)^k\, \Gamma(\alpha \!+\!1)/\Gamma(\alpha\!-\!k\!+\!1)$, the direct consequence of the previous formula
\[\fl 
\mbox{}\qquad 
{}_2F_1\left(\left.\!\!\!\!
\begin{array}{c}
-j\!-\!m,\, -j\!-\!n\\
1\!-\!n\!-\!m
\end{array}\!\!\right|\!-1\right)\!=\!\frac{\Gamma (1\!-\!n\!-\!m )\, \Gamma(2j\!+\!1)}{\Gamma (j\!-\!n\!+\!1)\, \Gamma(j\!-\!m\!+\!1)}\,
{}_2F_1\left(\left.\!\!\!\!
\begin{array}{c}
-j\!-\!m,\, -j\!-\!n\\
-2j
\end{array}\!\!\right|\!2\right).
\]
can be written as
\[
\begin{array}{l}
K_m(n)\!=\!C_{2j}^{j+m}\,
{}_2F_1\!\!\left(\left.\!\!\!\!
\begin{array}{c}
-j\!-\!m,\, -j\!-\!n\\
-2j
\end{array}\!\!\right|\!2\right).\qquad \opensquare
\end{array}
\]
In $\ell^2[-j,j]$, by admitting that
\[
K_m=0=\mathfrak{K}_m\qquad \mbox{for}\quad m\!\in\!\mathbb{Z}\backslash\{-j,-j\!+\!1,...,j\!-\!1,j\}
\]
we have
\[
\mathfrak{K}_m(n)=\mathfrak{K}_n(m),\qquad \mbox{for any}\ \ n,m\!\in \!\mathbb{Z}.
\]
{\bf Theorem 3}. {\it In $\ell^2[-j,j]$, the Kravchuk function $\mathfrak{K}_{n}$ satisfies the relation}
\begin{equation}\label{recK2}\fl
\mbox{}\qquad 
\sqrt{(j\!-\!m)(j\!+\!m\!+\!1)}\mathfrak{K}_{n}(m\!+\!1)\!+\!\sqrt{(j\!+\!m)(j\!-\!m\!+\!1)}\mathfrak{K}_{n}(m\!-\!1)\!=\!-2n\, \mathfrak{K}_n(m).
\end{equation}
{\it Proof}. By differentiating (\ref{defKpoly}) we get the relation
\[
\sum_{m=-j+1}^{j}\!\!\!(j\!+\!m)\, K_m(n)\, X^{j+m-1}\!+\!(2n\!+\!2jX)(1\!-\!X)^{j+n-1}(1\!+\!X)^{j-n-1}\!=\!0
\]
that is, the polynomial equality
\[\fl
(1\!-\!X)(1\!+\!X)\!\!\!\sum_{m=-j+1}^{j}\!\!\!(j\!+\!m)\, K_m(n)\, X^{j+m-1}\!+\!(2n\!+\!2jX)\!\!\!\sum_{m=-j}^{j}\!\!K_m(n)\, X^{j+m}\!=\!0
\]
leading to
\begin{equation}\label{recK}
(j\!+\!m\!+\!1)\, K_{m+1}(n)\!+\!(j\!-\!m\!+\!1)\, K_{m-1}(n)\!=\!-2n\, K_{m}(n).
\end{equation}
The corresponding recurrence relation for the Kravchuk functions is 
\begin{equation}\fl
\sqrt{(j\!-\!m)(j\!+\!m\!+\!1)}\ \mathfrak{K}_{m+1}(n)\!+\!\sqrt{(j\!+\!m)(j\!-\!m\!+\!1)}\ \mathfrak{K}_{m-1}(n)\!=\!-2n\, \mathfrak{K}_m(n).\qquad \opensquare
\end{equation}

The operators $J_z,\, J_+,\, J_-:\ell^2[-j,j]\longrightarrow \ell^2[-j,j]$  defined as \cite{Perelomov}
\begin{equation}
\begin{array}{l}
J_z|j;m\rangle =m\, |j;m\rangle \\[1mm]
J_+|j;m\rangle =\sqrt{(j\!-\!m)(j\!+\!m\!+\!1)}\ |j;m\!+\!1\rangle \\[1mm]
J_-|j;m\rangle =\sqrt{(j\!+\!m)(j\!-\!m\!+\!1)}\ |j;m\!-\!1\rangle 
\end{array}
\end{equation}
satisfy the relations
\begin{equation}
[J_z,J_{\pm}]=\pm J_{\pm},\qquad [J_-,J_+]=-2J_z
\end{equation}
and define an irreducible representation of the group $SU(2)$.  
The operators
\begin{equation}
J_x\!=\!\frac{1}{2}(J_+\!+J_-),\qquad J_y\!=\!\frac{1}{2{\rm i}}(J_+\!-\!J_-)\quad\mbox{and}\quad  J_z,
\end{equation}
satisfying the relations
\begin{equation}
[J_x,J_y]={\rm i}J_z,\qquad [J_y,J_z]={\rm i}J_x,\qquad [J_z,J_x]={\rm i}J_y,
\end{equation}
define an irreducible representation of the rotation group $SO(3)$. The operator corresponding to the rotation of angle $\omega$ around a unit vector $a=(a_1,a_2,a_3)\!\in \!\mathbb{R}^3$ is 
${\rm e}^{{\rm i}\omega(a_1J_x+a_2J_y+a_3J_z)}$. 
By using the relation
 \[
 \mathfrak{K}_n(m)\!=\!\langle j;m|\mathfrak{K}_n\rangle\!=\!\langle \mathfrak{K}_n|j;m\rangle.
 \]
 the equality (\ref{recK2}) can be written in the form
 \[
\langle \mathfrak{K}_n|J_+|j;m\rangle
 +\langle \mathfrak{K}_n|J_-|j;m\rangle\!=\!-2n\, \langle \mathfrak{K}_n|j;m\rangle 
 \]
 equivalent to
 \begin{equation}
 J_x|\mathfrak{K}_{-n}\rangle =n\, |\mathfrak{K}_{-n}\rangle .
 \end{equation}
 
The {\em Kravchuk transform}  \  $K:\ell^2[-j,j]\longrightarrow \ell^2[-j,j]$,
\begin{equation}
K=\sum_{n,m=-j}^j\mathfrak{K}_{-n}(m)\ |j;m\rangle\langle j;n| 
\end{equation}
is a unitary transform such that $K|j;n\rangle =|\mathfrak{K}_{-n}\rangle $.
It also satisfies the relations 
\begin{equation}
K^2\!=\!\sum_{n=-j}^j(-1)^{j+n}\, |j;-n\rangle \langle j;n|,\qquad K^4\!=\!\mathbb{I}\quad \mbox{and}\quad K^{-1}\!=\!K^3\!=\!K^+.
\end{equation}
We have
\begin{equation}
J_x=\sum_{n=-j}^j n\, |\mathfrak{K}_{-n}\rangle \langle \mathfrak{K}_{-n}|=
 \sum_{n=-j}^j n\, K|j;n\rangle \langle j;n|K^+
 = KJ_zK^+,
\end{equation}
but $K$ is not the only unitary operator satisfying this relation. The generalized Kravchuk transform \  $U:\ell^2[-j,j]\longrightarrow \ell^2[-j,j]$,
\begin{equation}
U=\sum_{n,m=-j}^j {\rm e}^{{\rm i}\alpha_n}\,\mathfrak{K}_{-n}(m)\ |j;m\rangle\langle j;n| 
\end{equation}
satisfies for any real numbers $\alpha _n$ the relation
\begin{equation}  
J_x=UJ_zU^+.
\end{equation}

A finite system $\{ |w_i\rangle \}_{i\in I}$ formed by $D$  non-null vectors from $\mathbb{C}^d$ is a {\em frame} if there exist two constants $0\!<\alpha \!\leq \!\beta \!<\infty $ such that
\begin{equation}\label{frame}
\alpha ||\psi ||^2\leq \sum_{i\in I} |\langle w_i|\psi \rangle |^2\leq \beta ||\psi ||^2\qquad \mbox{for all}\quad  |\psi \rangle \!\in \!\mathbb{C}^d.
\end{equation}
By using the frame operator $S=\sum_{i\in I} |w_i\rangle \langle w_i|$ this relation can be written in the form
\[
\langle \psi|\alpha \, \psi \rangle \leq \langle \psi|S \psi \rangle  \leq \langle \psi|\beta \, \psi \rangle \qquad \mbox{for all}\quad  |\psi \rangle \!\in \!\mathbb{C}^d,  
\]
that is \  $\alpha  \mathbb{I}\!\leq \!S\!\leq \!\beta  \mathbb{I}.$ A frame with $\alpha\!=\!\beta\!=\!1$,  called a {\em tight frame}, satisfies the relation
\begin{equation}\label{frame-id}
\sum_{i\in I} |w_i\rangle \langle w_i|=\mathbb{I}.
\end{equation}
By denoting $\kappa _i\!=\!\langle w_i|w_i\rangle $ and $|u_i\rangle \!=\!\frac{1}{\sqrt{\kappa _i}}|w_i\rangle $,
the resolution of the identity (\ref{frame-id}) becomes
\begin{equation}\label{frame-idd}
\sum_{i\in I} \kappa _i\, |u_i\rangle \langle u_i|=\mathbb{I}.
\end{equation}
We have
\begin{equation}
\langle \psi|\varphi\rangle=\sum_{i\in I} \kappa _i\langle\psi|u_i\rangle \langle u_i|\varphi\rangle ,\qquad 
||\psi ||^2\!=\!\langle \psi|\psi\rangle=\sum_{i\in I} \kappa _i\, |\langle u_i|\psi\rangle |^2
\end{equation}
and 
\[
\begin{array}{l}
d\!=\!\sum\limits_{n=-j}^j\langle j;n|j;n\rangle\!=\!\sum\limits_{n=-j}^j\sum\limits_{i\in I}\kappa _i\, |\langle u_i|j;n\rangle |^2\!=\!\sum\limits_{i\in I}\kappa _i .
\end{array}
\]
Particularly, if all  $\kappa _i$ are equal, then  
\begin{equation}
\begin{array}{l}
\frac{d}{D}\sum\limits_{i\in I} |u_i\rangle \langle u_i|=\mathbb{I}\qquad \mbox{and}\qquad 
\langle \psi|\varphi\rangle=\frac{d}{D}\sum\limits_{i\in I} \langle\psi|u_i\rangle \langle u_i|\varphi\rangle .
\end{array}
\end{equation}
Each frame $\{|u_i\rangle \}_{i\in I}$ satisfying (\ref{frame-idd})  defines an embedding of $\mathbb{C}^d$ into a larger space $\mathbb{C}^D$.
In the Hilbert space $\mathbb{C}^D$, the vectors 
\[
|v_n\rangle =\sum_{i\in I} \sqrt{\kappa _i} \, \langle u_i|j;n\rangle \, |e_i\rangle \qquad \mbox{with}\qquad n\!\in \!\{-j,-j\!+\!1, ...,j\!-\!1,j\},
\]
defined by using an orthonormal basis $\{|e_i\rangle \}_{i\in I}$ in $\mathbb{C}^D$,
form an orthonormal system.\\
Our Hilbert space $\mathbb{C}^d$ can be identified with the subspace
\[
\mathcal{H}={\rm span}\{|v_{-j}\rangle ,\, |v_{-j+1}\rangle ,\, ...,\,|v_{j-1}\rangle ,\,|v_{j}\rangle \}
\]
by using the isometry
\begin{equation}
\begin{array}{l}
\mathbb{C}^d\longrightarrow \mathcal{H}:\ \sum\limits_{n=-j}^j 
\alpha_n\, |j;n\rangle \ \mapsto \sum\limits_{n=-j}^j 
\alpha_n\, |v_n\rangle .
\end{array}
\end{equation}
After identification, $|u_i\rangle $ satisfies the relation
\begin{equation}
\begin{array}{l}
|u_i\rangle \equiv \sum\limits_{n=-j}^j |v_n\rangle\langle j;n|u_i\rangle 
=\frac{1}{\sqrt{\kappa _i}}\sum\limits_{n=-j}^j  |v_n\rangle\langle v_n|e_i\rangle =\frac{1}{\sqrt{\kappa _i}}\, \mathcal{P}|e_i\rangle
,
\end{array}
\end{equation}
where $\mathcal{P}$ is the orthogonal projector 
\begin{equation}
\begin{array}{l}
\mathcal{P}\!=\!\sum_{n=-j}^j  |v_n\rangle\langle v_n|\qquad \mbox{and}\qquad \kappa _i\!=\!\langle e_i|\mathcal{P}|e_i\rangle.
\end{array}
\end{equation}
Thus, the frame  $\{|u_i\rangle \}_{i\in I}$ is the normalized projection of the orthonormal basis $\{|e_i\rangle \}_{i\in I}$.\\
By using the frame $\{|u_i\rangle \}_{i\in I}$, we can associate to a function
$f:I\longrightarrow \mathbb{C}$ the operator
\begin{equation}
A_f:\mathbb{C}^d\longrightarrow \mathbb{C}^d,\qquad A_f=\sum_{i\in I} \kappa _i\,f(i)\,  |u_i\rangle \langle u_i|
\end{equation}
and to a linear operator $A:\mathbb{C}^d\longrightarrow \mathbb{C}^d$ the function
\begin{equation}
f_A:I\longrightarrow \mathbb{C},\qquad  f_A(i)=\langle u_i|A|u_i\rangle .
\end{equation}
Some useful mathematical objects can be defined by using the quantization 
$f\mapsto A_f$ and the dequantization $A\mapsto f_A$ procedures.

\begin{figure}
\centering
\includegraphics[scale=1.1]{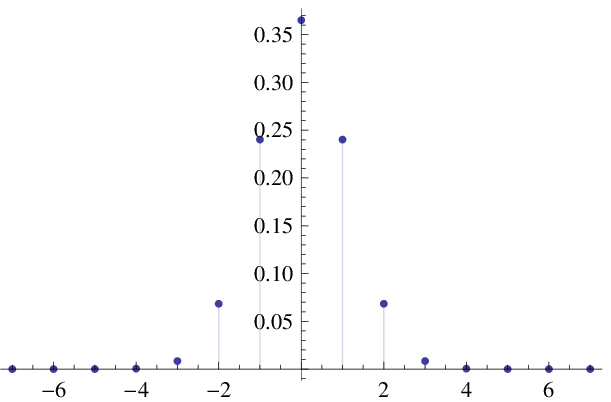}\ \ 
\includegraphics[scale=1.0]{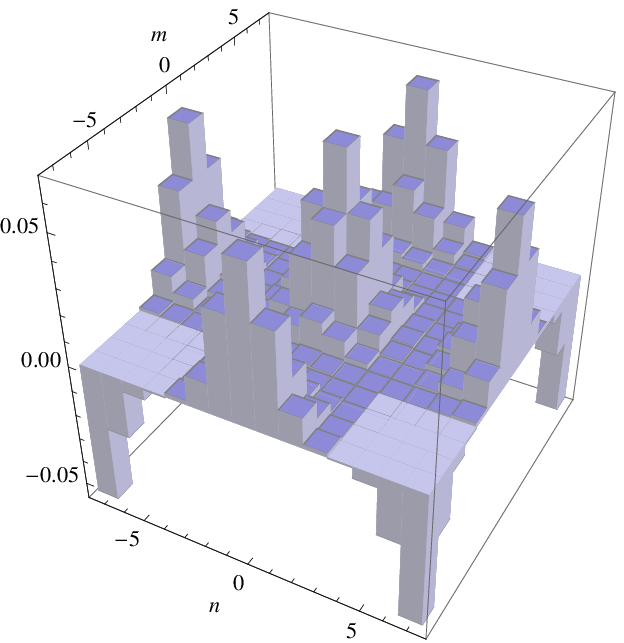}
\caption{\label{gW} The probability distribution $|\mathfrak{G}_1|^2$ corresponding to $\mathfrak{G}_1=\mathfrak{g}_1^{(1)}/||\mathfrak{g}_1^{(1)}||$ and the discrete Wigner function $W_{\mathfrak{G}_1}=W_{\mathfrak{g}_1^{(1)}}/||\mathfrak{g}_1^{(1)}||^2$ defined by (\ref{Wg}), in the case $d\!=\!15$.}
\end{figure}

\section{Finite Gaussians}

The Gaussian functions of the form $a g^{(\kappa)}$, where
\begin{equation} 
g^{(\kappa)} :\mathbb{R}\longrightarrow \mathbb{R},\qquad 
g^{(\kappa)}(q)={\rm e}^{-\frac{\kappa }{2}q^2},
\end{equation}
and $a,\, \kappa \!\in \!(0,\infty )$ are parameters, play an important role in quantum mechanics, signal processing and mathematics.
They belong to the Hilbert space $L^2(\mathbb{R})$ of the square integrable functions, and the corresponding norm can be evaluated exactly
\begin{equation}
||g^{(\kappa)} ||^2=\int_{-\infty }^\infty |g^{(\kappa)}(q)|^2\, dq=
\int_{-\infty }^\infty {\rm e}^{-\kappa q^2}\, dq=\sqrt{\frac{\pi }{\kappa }}.
\end{equation}

The direct/inverse Fourier transform of a function $f:\mathbb{R}\longrightarrow \mathbb{C}$ is the function 
\begin{equation}
\mathcal{F}^{\pm 1}[f]:\mathbb{R}\longrightarrow \mathbb{C},\qquad
\mathcal{F}^{\pm 1}[f](p)=\frac{1}{\sqrt{2\pi }}\int_{-\infty }^\infty {\rm e}^{\mp {\rm i}pq}\, f(q)\, dq
\end{equation}
and the corresponding Wigner function  is the function 
\begin{equation}
W_{\!f}:\mathbb{R}^2\longrightarrow \mathbb{R},\qquad 
W_{\!f}(q,p) =\frac{1}{\pi}\int_{-\infty}^\infty  \mathrm{e}^{2{\rm i}pu }f \left(q\!-\!u\right)\, \overline{f\left(q\!+\!u\right)} \, du.
\end{equation}
The Fourier transform of a Gaussian function is also a Gaussian function 
\begin{equation}
\begin{array}{l}
\mathcal{F}[g^{(\kappa)}]=\frac{1}{\sqrt{\kappa }}\, g^{(1/\kappa)}
\end{array}
\end{equation}
and the corresponding Wigner function  is a product of Gaussian functions
\begin{equation}
\begin{array}{l}
W_{\!g^{(\kappa)}}(q,p)=\frac{1}{\sqrt{\kappa \pi}}\, g^{(2\kappa)} (q)\,g^{(2/\kappa)}(p).
\end{array}
\end{equation}

In this section, our aim is to investigate the even functions (see Fig. 1-4)
\[
\mathfrak{g}_1^{(\kappa)},\, 
\mathfrak{g}_2^{(\kappa)},\,  \mathfrak{g}_3^{(\kappa)},\, \mathfrak{g}_4,\, \mathfrak{g}_5\!:\!\{-j,\ -j\!+\!1,\ ...\ ,\ j\!-\!1,\ j\}\!\longrightarrow \!\mathbb{R},
\]
\begin{equation}\label{defg}
\begin{array}{l}
\mathfrak{g}_1^{(\kappa)}(n)=\sum\limits_{\alpha =-\infty}^{\infty}{\rm e}^{-\frac{\kappa\pi}{d}(\alpha d+n)^2},
\end{array}
\end{equation}
\begin{equation}\label{defgp}
\begin{array}{l}
\mathfrak{g}_2^{(\kappa)}(n)=\sum\limits_{\alpha =-\infty}^{\infty}{\rm e}^{-\frac{\kappa\pi}{d}((\alpha +\frac{1}{2})d+n)^2},
\end{array}
\end{equation}
\begin{equation}\label{defgm}
\begin{array}{l}
\mathfrak{g}_3^{(\kappa)}(n)=(-1)^n\sum\limits_{\alpha =-\infty}^{\infty}(-1)^\alpha \, {\rm e}^{-\frac{\kappa\pi}{d}(\alpha d+n)^2},
\end{array}
\end{equation}
\begin{equation}\label{defb}
\begin{array}{l}
\mathfrak{g}_4(n)=\frac{1}{2^{2j}}\frac{(2j)!}{(j\!-\!n)!\, (j\!+\!n)!},
\end{array}
\end{equation}
\begin{equation}\label{defc}
\begin{array}{l}
\mathfrak{g}_5(n)=\frac{1}{\sqrt{d}}\, \cos^{2j}\frac{n\pi }{d},
\end{array}
\end{equation}
 where $\kappa \!\in \!(0,\infty)$ is a parameter. We call {\em finite Gaussians} the functions of the form
 \[
a\mathfrak{g}_1^{(\kappa)},\, 
a\mathfrak{g}_2^{(\kappa)},\,  a\mathfrak{g}_3^{(\kappa)},\, a\mathfrak{g}_4,\, a\mathfrak{g}_5\qquad \mbox{with}\quad a\!\in \!(0,\infty).
 \]
\begin{figure}
\centering
\includegraphics[scale=1.1]{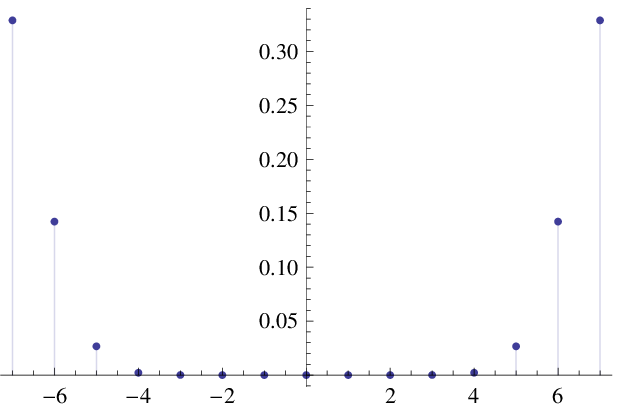}\ \ 
\includegraphics[scale=1.0]{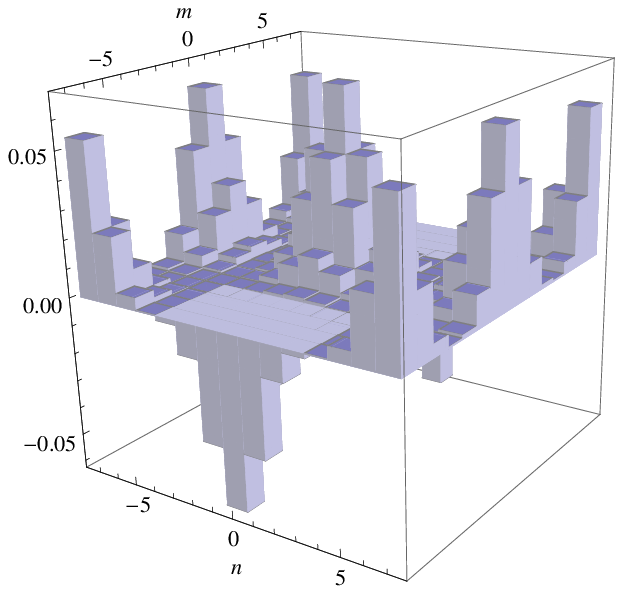}
\caption{\label{gpW} The probability distribution $|\mathfrak{G}_2|^2$ corresponding to $\mathfrak{G}_2=\mathfrak{g}_2^{(1)}/||\mathfrak{g}_2^{(1)}||$ and the discrete Wigner function $W_{\mathfrak{G}_2}=W_{\mathfrak{g}_2^{(1)}}/||\mathfrak{g}_2^{(1)}||^2$ defined by (\ref{Wgp}), in the case $d\!=\!15$.}
\end{figure}
\noindent{\bf Theorem 4}. {\em 
The Fourier transform of a finite Gaussian is a finite Gaussian:}
\begin{equation}\label{Fgk}
\begin{array}{l}
a)\quad \qquad \qquad\qquad \  F[\mathfrak{g}_1^{(\kappa)}]=\frac{1}{\sqrt{\kappa}}\, \mathfrak{g}_1^{(1/\kappa)},
\end{array}
\end{equation}
\begin{equation}\label{Fgkpm}
\begin{array}{l}
b)\quad F[\mathfrak{g}_2^{(\kappa)}]\!=\!\frac{1}{\sqrt{\kappa}}\, \mathfrak{g}_3^{(1/\kappa)}\quad \quad  and \qquad \qquad F[\mathfrak{g}_3^{(\kappa)}]=\frac{1}{\sqrt{\kappa}}\, \mathfrak{g}_2^{(1/\kappa)},
\end{array}
\end{equation}
\begin{equation}\label{Fbc}
\begin{array}{l}
c)\quad \ \ \ F[\mathfrak{g}_4]=\mathfrak{g}_5\qquad \quad\quad  and \qquad \qquad \quad  F[\mathfrak{g}_5]=\mathfrak{g}_4.
\end{array}
\end{equation}
{\it Proof}. a) A proof can be found in \cite{CD}.\\
b) 
The periodic function $G_2^{(\kappa)}:\mathbb{R}\longrightarrow \mathbb{R}$
\[ 
\begin{array}{l}
G_2^{(\kappa)}(x)=\sum\limits_{\alpha =-\infty }^\infty \mathrm{e}^{-\frac{\kappa \pi }{d} \left(\left(\alpha + \frac{1}{2}\right)d +x\right)^2}=\sum\limits_{\alpha =-\infty }^\infty \mathrm{e}^{-\frac{\kappa }{2}\left(\sqrt{\frac{2\pi }{d}}\, \left(\left(\alpha + \frac{1}{2}\right)d +x\right)\right)^2}
\end{array}
\]
with period $d$ admits the Fourier expansion\\[-3mm]
\[
\begin{array}{l}
G_2^{(\kappa)}(x)=\sum\limits_{\ell =-\infty }^\infty c_\ell \,  \mathrm{e}^{\frac{2\pi \mathrm{i}}{d}\ell x}
\end{array}
\]
with\\[-3mm]
\[
\begin{array}{rl}
c_\ell & =\frac{1}{d}\int _0^d\mathrm{e}^{-\frac{2\pi \mathrm{i}}{d}\ell x}\sum\limits_{\alpha =-\infty }^\infty \mathrm{e}^{-\frac{\kappa}{2} \left(\sqrt{\frac{2\pi }{d}}\, \left(\left(\alpha + \frac{1}{2}\right)d +x\right)\right)^2} dx\\[2mm]
& =\frac{1}{d}\sum\limits_{\alpha =-\infty }^\infty \int _0^d\mathrm{e}^{-\frac{2\pi \mathrm{i}}{d}\ell x} \mathrm{e}^{-\frac{\kappa}{2}\left(\sqrt{\frac{2\pi }{d}}\, \left(\left(\alpha + \frac{1}{2}\right)d +x\right)\right)^2} dx.
\\[-1mm]
\end{array}
\]
By denoting $t\!=\!\sqrt{\frac{2\pi }{d}}\, \left(\left(\alpha + \frac{1}{2}\right)d +x\right)$ 
we get \cite{Mehta}
\[
\begin{array}{rl}
c_\ell & =\frac{1}{\sqrt{2\pi d}}\sum\limits_{\alpha =-\infty }^\infty \int_{(\alpha -1/2)\sqrt{2\pi d}}^{(\alpha +1/2)\sqrt{2\pi d}}\mathrm{e}^{-\frac{2\pi \mathrm{i}}{d}\ell \left(t\sqrt{\frac{d}{2\pi }}-\left(\alpha +\frac{1}{2}\right) d\right)} \mathrm{e}^{-\frac{\kappa}{2} t^2}\, dt\\[4mm]
 & =\frac{(-1)^\ell }{\sqrt{2\pi d}}\sum\limits_{\alpha =-\infty }^\infty \int_{(\alpha -1/2)\sqrt{2\pi d}}^{(\alpha +1/2)\sqrt{2\pi d}}\mathrm{e}^{-\mathrm{i}\ell  t\sqrt{\frac{2\pi }{d}}}
\mathrm{e}^{-\frac{\kappa}{2}t^2}\, dt\\[4mm]
& =\frac{(-1)^\ell }{\sqrt{2\pi d}}\int_{-\infty }^{\infty}\mathrm{e}^{-\mathrm{i}\ell  t\sqrt{\frac{2\pi }{d}}} \mathrm{e}^{-\frac{\kappa}{2} t^2}\,  dt=\frac{(-1)^\ell }{\sqrt{\kappa d}}\, \mathrm{e}^{-\frac{\pi}{\kappa d}\ell ^2}
\end{array}
\]
whence
\[
\begin{array}{l}
G_2^{(\kappa)}(x)=\frac{1}{\sqrt{\kappa d}}\,\sum\limits_{\ell =-\infty }^\infty   \mathrm{e}^{\frac{2\pi \mathrm{i}}{d}\ell x}\,(-1)^\ell \,  \mathrm{e}^{-\frac{\pi}{\kappa d}\ell ^2}.
\end{array}
\]
Particularly, we have \cite{Mehta}
\[
\begin{array}{rl}
\mathfrak{g}_2^{(\kappa)}(k) & \!\!\!\!=G_2^{(\kappa)}(k)=\frac{1}{\sqrt{\kappa d}}\,\sum\limits_{\ell =-\infty }^\infty   \mathrm{e}^{\frac{2\pi \mathrm{i}}{d}k\ell }\,(-1)^\ell \, \mathrm{e}^{-\frac{\pi}{\kappa d}\ell ^2}\\[4mm]
 & \!\!\!\!=\frac{1}{\sqrt{\kappa d}}\sum\limits_{n=-s}^{s}\sum\limits_{\alpha =-\infty }^\infty  \mathrm{e}^{\frac{2\pi \mathrm{i}}{d}k(\alpha d+n) }\,(-1)^{(\alpha d+n)} \mathrm{e}^{-\frac{\pi }{\kappa d}(\alpha d+n) ^2}\, \\[4mm]
& \!\!\!\!=\!\frac{1}{\sqrt{d}}\sum\limits_{n=-s}^{s}\!\!\mathrm{e}^{\frac{2\pi \mathrm{i}}{d}kn }\frac{(-1)^n}{\sqrt{\kappa }}\!\!\sum\limits_{\alpha =-\infty }^\infty  \!\!(-1)^\alpha \,  \mathrm{e}^{-\frac{\pi }{\kappa d}(\alpha d+n) ^2}
\!\!=\!\frac{1}{\sqrt{\kappa }}F[\mathfrak{g}_3^{(1/\kappa)}](k).
\end{array}
\]
c) We have
\[
\begin{array}{rl}
F[\mathfrak{g}_4](k) & \!\!\!\!=\frac{1}{\sqrt{d}}\frac{1}{2^{2j}}\sum\limits_{n=-j}^j \frac{(2j)!}{(j\!-\!n)!\, (j\!+\!n)!} {\rm e}^{-\frac{2\pi {\rm i}}{d}kn}\\[3mm]
& \!\!\!\!=\frac{1}{\sqrt{d}}\frac{1}{2^{2j}} 
\left({\rm e}^{\frac{\pi {\rm i}}{d}kj}\!+\!{\rm e}^{-\frac{\pi {\rm i}}{d}k}\right)^{2j}=\frac{1}{\sqrt{d}}\, \cos^{2j}\frac{k\pi }{d}=\mathfrak{g}_5(k).\qquad \opensquare    
\end{array}
\]
In this paper, we are mainly interested in the normalized Gaussians (see Fig. 1-4)
\begin{equation}\fl
\mathfrak{G}_1\!=\!\frac{\mathfrak{g}_1^{(1)}}{||\mathfrak{g}_1^{(1)}||},\quad 
\mathfrak{G}_2\!=\!\frac{\mathfrak{g}_2^{(1)}}{||\mathfrak{g}_2^{(1)}||},\quad 
\mathfrak{G}_3\!=\!\frac{\mathfrak{g}_3^{(1)}}{||\mathfrak{g}_3^{(1)}||},\quad 
\mathfrak{G}_4\!=\!\frac{\mathfrak{g}_4}{||\mathfrak{g}_4||},\quad 
\mathfrak{G}_5\!=\!\frac{\mathfrak{g}_5}{||\mathfrak{g}_5||}
\end{equation}
satisfying the relations
\begin{equation}\fl
F[\mathfrak{G}_1]\!=\!\mathfrak{G}_1,\quad F[\mathfrak{G}_2]\!=\!\mathfrak{G}_3,\quad F[\mathfrak{G}_3]\!=\!\mathfrak{G}_2,\quad
F[\mathfrak{G}_4]\!=\!\mathfrak{G}_5,\quad F[\mathfrak{G}_5]\!=\!\mathfrak{G}_4.
\end{equation}
Due to the periodicity, the distribution of probability $|\mathfrak{G}_2|^2$ corresponding to $\mathfrak{G}_2$ has the shape of a Gaussian function (see Fig. \ref{gpW}). In this case the maximum value is taken in the neighbouring points $-j$ and $j$.
The function $\mathfrak{G}_3$ takes positive as well as negative values,
but the corresponding distribution of probability $|\mathfrak{G}_3|^2$ has the shape of a Gaussian function (see Fig. \ref{gmW}). 

\begin{figure}
\centering
\includegraphics[scale=1.1]{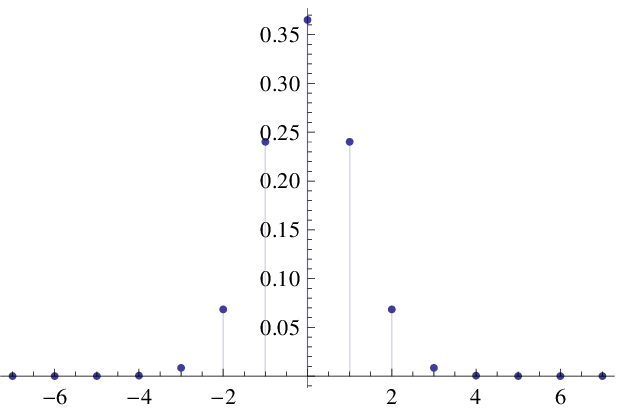}\ \
\includegraphics[scale=1.1]{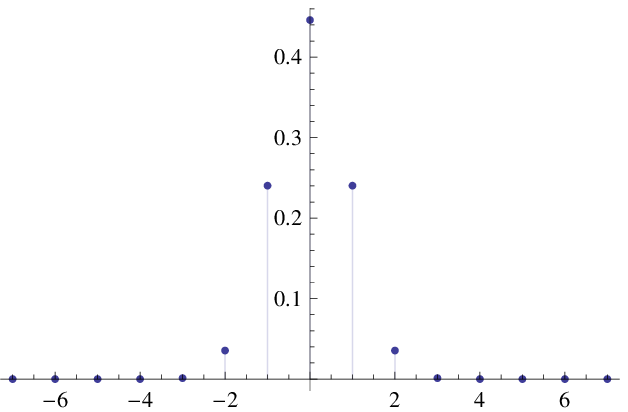}
\caption{\label{gmW} The probability distribution $|\mathfrak{G}_3|^2$ corresponding to $\mathfrak{G}_3=\mathfrak{g}_3^{(1)}/||\mathfrak{g}_3^{(1)}||$ and the probability distribution $|\mathfrak{G}_5|^2$ corresponding to $\mathfrak{G}_5=\mathfrak{g}_5^{(1)}/||\mathfrak{g}_5^{(1)}||$, in the case $d\!=\!15$.}
\end{figure}

The Jacobi theta function \cite{Magnus,Vilenkin,Whittaker}
\begin{equation} \label{theta}
\begin{array}{l}
\theta _3(z,\tau )=\sum\limits_{\alpha =-\infty }^\infty {\rm e}^{{\rm i}\pi \tau \alpha ^2}\, {\rm e}^{2\pi {\rm i}\alpha z}
\end{array}
\end{equation}
has several remarkable properties among which we mention:
\[
\theta _3(z+m+n\tau ,\tau )={\rm e}^{-{\rm i}\pi \tau n^2}\, {\rm e}^{-2\pi {\rm i}nz}\, \theta _3(z,\tau )
\]
\begin{equation} \label{theta3prop}
\begin{array}{l}
\theta _3(z,{\rm i}\tau )=\frac{1}{\sqrt{\tau }}\, \mathrm{e}^{-\frac{\pi z^2}{\tau }}\, \theta _3 \left( \frac{z}{{\rm i}\tau },\frac{\rm i}{\tau }\right)
\end{array}
\end{equation}
and \cite{MR,Ruzzi}
\begin{equation} \label{Ruzzi}
\begin{array}{l}
\theta _3 \left( \frac{k}{d},\frac{{\rm i}\kappa }{d}\right)=\frac{1}{\sqrt{\kappa d}}\, \sum\limits_{n=-s}^{s}{\rm e}^{-\frac{2\pi {\rm i}}{d}kn}\, \theta _3 \left( \frac{n}{d},\frac{\rm i}{ \kappa d}\right).
\end{array}
\end{equation}
By using (\ref{theta3prop}) we get
\[\begin{array}{l}
\theta _3\left( \frac{n}{d},\frac{{\rm i}}{\kappa d} \right)=\sqrt{\kappa d}\ \mathrm{e}^{-\frac{\kappa \pi }{d} n^2} \theta _3\left( -{\rm i}\kappa n, {\rm i}\kappa d\right)=\sqrt{\kappa d}\sum\limits_{\alpha =-\infty }^\infty \mathrm{e}^{-\frac{\kappa \pi }{d} (\alpha d +n)^2}
\end{array}
\]
that is, the formula
\begin{equation}
\begin{array}{l}
\mathfrak{g}_1^{(\kappa)} (n)=\frac{1}{\sqrt{\kappa d}}\, \theta _3\left( \frac{n}{d},\frac{{\rm i}}{\kappa d} \right)
\end{array}
\end{equation}
Ruzzi's relation (\ref{Ruzzi}) is equivalent to our formula (\ref{Fgk}).

By using the Jacobi function
\begin{equation}
\begin{array}{l}
\theta _4(z,\tau )=\theta _3\left(z\!+\!\frac{1}{2},\tau \right)=\sum\limits_{\alpha =-\infty }^\infty (-1)^\alpha \ {\rm e}^{{\rm i}\pi \tau \alpha ^2}\, {\rm e}^{2\pi {\rm i}\alpha z}
\end{array}
\end{equation}
and (\ref{theta3prop}) we get the relation
\[\begin{array}{rl}
\theta _4\left( \frac{n}{d},\frac{{\rm i}}{\kappa d} \right) & \!\!\!\!=
\theta _4\left( \frac{n}{d}+\frac{1}{2},\frac{{\rm i}}{\kappa d} \right)\\[2mm]
& \!\!\!\!=\!
\sqrt{\kappa d}\ \mathrm{e}^{-\frac{\kappa \pi }{d} \left(n+\frac{d}{2}\right)^2} \theta _3\left( -{\rm i}\kappa \left(n+\frac{d}{2}\right), {\rm i}\kappa d\right)\\[2mm]
 & \!\!\!\!=\sqrt{\kappa d}\sum\limits_{\alpha =-\infty }^\infty \mathrm{e}^{-\frac{\kappa \pi }{d} \left(\left(\alpha +\frac{1}{2}\right)d +n\right)^2}
\end{array}
\]
equivalent to
\begin{equation} \label{gkpJ}
\begin{array}{l}
\mathfrak{g}_2^{(\kappa)}(n)=\frac{1}{\sqrt{\kappa d}}\, \theta _4\left( \frac{n}{d},\frac{{\rm i}}{\kappa d} \right).
\end{array}
\end{equation}
In a similar way, by using the Jacobi function
\[\begin{array}{l}
\theta _2(z,\tau )=\mathrm{e}^{\frac{1}{4}\pi {\rm i}\tau+\pi {\rm i}z}\, \theta _3\left(z\!+\!\frac{\tau}{2},\tau \right)=\sum\limits_{\alpha =-\infty }^\infty {\rm e}^{{\rm i}\pi \tau \left(\alpha +\frac{1}{2}\right)^2}\, {\rm e}^{2\pi {\rm i}\left(\alpha +\frac{1}{2}\right)z}
\end{array}
\]
and (\ref{theta3prop}) we obtain the formula
\[\begin{array}{rl}
\theta _2\left( \frac{n}{d},\frac{{\rm i}}{\kappa d} \right)& \!\!\!\!=\mathrm{e}^{-\frac{\pi}{4\kappa d}+\frac{\pi {\rm i}}{d}n}\ 
\theta _3\left( \frac{n}{d}\!+\!\frac{{\rm i}}{2\kappa d},\frac{{\rm i}}{\kappa d} \right)\\[2mm]
& \!\!\!\!=\!
\sqrt{\kappa d}\ \mathrm{e}^{-\frac{\pi}{4\kappa d}+\frac{\pi {\rm i}}{d}n}\ \mathrm{e}^{-\frac{\kappa \pi }{d} \left(n+\frac{\rm i}{2\kappa}\right)^2} \theta _3\left( -{\rm i}\kappa n+\frac{1}{2}, {\rm i}\kappa d\right)\\[2mm]
 & \!\!\!\!=\sqrt{\kappa d}\sum\limits_{\alpha =-\infty}^{\infty}(-1)^\alpha {\rm e}^{-\frac{\kappa\pi}{d}(\alpha d+n)^2}
\end{array}
\]
equivalent to
\begin{equation} \label{gkmJ}
\begin{array}{l}
\mathfrak{g}_3^{(\kappa)}(n)=\frac{(-1)^n}{\sqrt{\kappa d}}\, \theta _2\left( \frac{n}{d},\frac{{\rm i}}{\kappa d} \right).
\end{array}
\end{equation}

By writing
\[
\begin{array}{l}
\mathfrak{g}_1^{(\kappa)}(2m)=\sum\limits_{\alpha =-\infty}^{\infty}{\rm e}^{-\frac{\kappa\pi}{d}(2\alpha d+2m)^2}+\sum\limits_{\alpha =-\infty}^{\infty}{\rm e}^{-\frac{\kappa\pi}{d}[(2\alpha +1)d+2m]^2},\\[3mm]
\mathfrak{g}_3^{(\kappa)}(2m)\!=\!\sum\limits_{\alpha =-\infty}^{\infty}(-1)^{2\alpha} {\rm e}^{-\frac{\kappa\pi}{d}(2\alpha d+2m)^2}\!+\!\sum\limits_{\alpha =-\infty}^{\infty}(-1)^{2\alpha+1} {\rm e}^{-\frac{\kappa\pi}{d}[(2\alpha+1) d+2m]^2},
\end{array}
\]
we get the relations
\begin{equation}\label{g2n}
\begin{array}{l}
\mathfrak{g}_1^{(\kappa)}(2m)=\mathfrak{g}_1^{(4\kappa)}(m)+\mathfrak{g}_2^{(4\kappa)}(m),\\[2mm]
\mathfrak{g}_3^{(\kappa)}(2m)=\mathfrak{g}_1^{(4\kappa)}(m)-\mathfrak{g}_2^{(4\kappa)}(m).
\end{array}
\end{equation}

If we separate an absolutely convergent series as
\[
\sum\limits_{\alpha ,\beta =-\infty }^\infty N_{\alpha ,\beta }=\sum\limits_{\scriptsize 
\begin{array}{c}
\alpha ,\beta \\
{\rm both\ even}\\
{\rm or}\\
{\rm both\ odd}
\end{array}} N_{\alpha ,\beta }+\sum\limits_{\scriptsize 
\begin{array}{c}
\alpha ,\beta \\
{\rm one\ even}\\
{\rm and}\\
{\rm other\ odd}
\end{array}} N_{\alpha ,\beta }
\]
and use the substitutions $(\alpha , \beta )\!=\!(\mu \!+\!\eta ,\mu \!-\!\eta )$ and $(\alpha , \beta )\!=\!(\mu \!+\!\eta \!+\!1,\mu \!-\!\eta )$ we obtain 
\begin{equation}\label{doublesum}
\sum\limits_{\alpha ,\beta =-\infty }^\infty N_{\alpha ,\beta }=\sum\limits_{\mu  ,\eta =-\infty }^\infty N_{\mu +\eta ,\mu -\eta } + \sum\limits_{\mu  ,\eta  =-\infty }^\infty N_{\mu +\eta +1,\mu -\eta }.
\end{equation}
\begin{figure}
\centering
\includegraphics[scale=1.1]{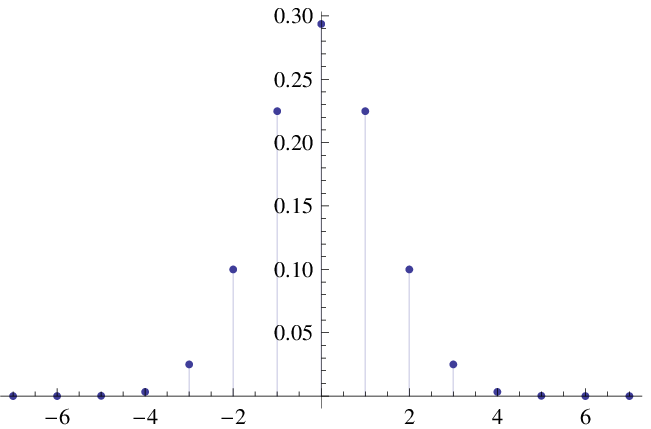}\ \
\includegraphics[scale=1.0]{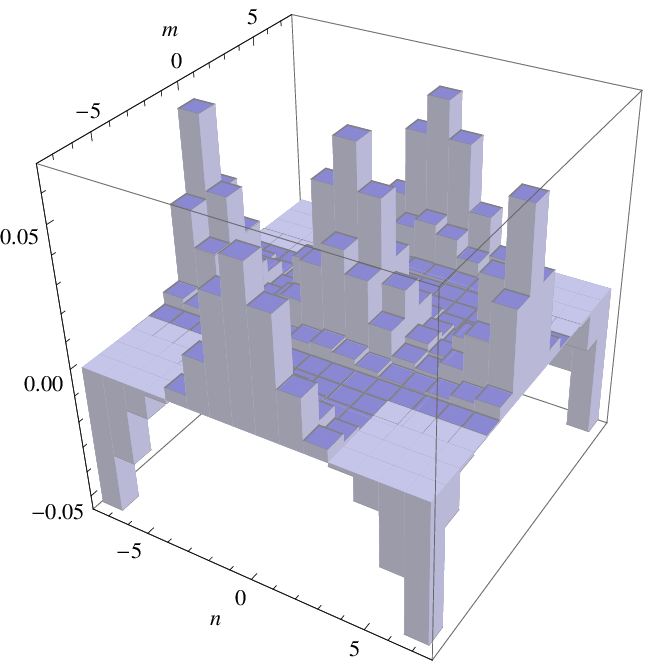}
\caption{\label{bW} The probability distribution $|\mathfrak{G}_4|^2$ corresponding to $\mathfrak{G}_4=\mathfrak{g}_4/||\mathfrak{g}_4||$ and the discrete Wigner function $W_{\mathfrak{G}_4}=W_{\mathfrak{g}_4}/||\mathfrak{g}_4||^2$, in the case $d\!=\!15$. }
\end{figure}
%
{\bf Theorem 5}. {\it The Wigner functions corresponding to $\mathfrak{g}_1^{(\kappa)}$, $\mathfrak{g}_2^{(\kappa)}$ and $\mathfrak{g}_3^{(\kappa)}$ are 
 sums\\
 \mbox{}\qquad \qquad \qquad  of products of finite Gaussians {\rm (see Fig. 1-2)}:}
\begin{equation}\label{Wg}
\begin{array}{rl} 
a)\quad W_{\mathfrak{g}_1^{(\kappa)}}(n,m) &\!\!\! \!=\!\frac{1}{\sqrt{2\kappa d}}\, \,  \mathfrak{g}_1^{(2\kappa)}(n)\,  \mathfrak{g}_1^{(2/\kappa)}(m)\!+\!\frac{1}{\sqrt{2\kappa d}}\, \,  \mathfrak{g}_1^{(2\kappa)}(n)\, \mathfrak{g}_2^{(2/\kappa)}(m)\\[3mm]
&  \ +\frac{1}{\sqrt{2\kappa d}}\,  \mathfrak{g}_2^{(2\kappa)}(n)\, \mathfrak{g}_1^{(2/\kappa)} (m)\!-\!\frac{1}{\sqrt{2\kappa d}}\,  \mathfrak{g}_2^{(2\kappa)}(n)\,\mathfrak{g}_2^{(2/\kappa)} (m),
\end{array}
\end{equation}
\begin{equation}\label{Wgp}
\begin{array}{rl} 
b)\quad W_{\mathfrak{g}_2^{(\kappa)}}(n,m) &\!\!\! \!=\!\frac{1}{\sqrt{2\kappa d}}\, \,  \mathfrak{g}_1^{(2\kappa)}(n)\,  \mathfrak{g}_1^{(2/\kappa)}(m)\!-\!\frac{1}{\sqrt{2\kappa d}}\, \,  \mathfrak{g}_1^{(2\kappa)}(n)\, \mathfrak{g}_2^{(2/\kappa)}(m)\\[3mm]
&  \ +\frac{1}{\sqrt{2\kappa d}}\,  \mathfrak{g}_2^{(2\kappa)}(n)\, \mathfrak{g}_1^{(2/\kappa)} (m)\!+\!\frac{1}{\sqrt{2\kappa d}}\,  \mathfrak{g}_2^{(2\kappa)}(n)\,\mathfrak{g}_2^{(2/\kappa)} (m),
\end{array}
\end{equation}
\begin{equation}\label{Wgm}
\begin{array}{rl} 
c)\quad W_{\mathfrak{g}_3^{(\kappa)}}(n,m) &\!\!\! \!=\!\frac{1}{\sqrt{2\kappa d}}\, \,  \mathfrak{g}_1^{(2\kappa)}(n)\,  \mathfrak{g}_1^{(2/\kappa)}(m)\!+\!\frac{1}{\sqrt{2\kappa d}}\, \,  \mathfrak{g}_1^{(2\kappa)}(n)\, \mathfrak{g}_2^{(2/\kappa)}(m)\\[3mm]
&  \ -\frac{1}{\sqrt{2\kappa d}}\,  \mathfrak{g}_2^{(2\kappa)}(n)\, \mathfrak{g}_1^{(2/\kappa)} (m)\!+\!\frac{1}{\sqrt{2\kappa d}}\,  \mathfrak{g}_2^{(2\kappa)}(n)\,\mathfrak{g}_2^{(2/\kappa)} (m).
\end{array}
\end{equation}
{\bf Proof}. a)  A proof can be found in \cite{CD}.\\
b) By using (\ref{Fgk}), (\ref{Fgkpm}), (\ref{g2n}) and (\ref{doublesum}) we get
\[ \fl
\begin{array}{rl}
W_{\mathfrak{g}_2^{(\kappa)}}(n,m)&\!\!\!\!=\frac{1}{d}\sum\limits_{k=-s}^s\mathrm{e}^{\frac{4\pi \mathrm{i}}{d}mk}\, 
\sum\limits_{\mu  ,\eta  =-\infty }^\infty \mathrm{e}^{-\kappa  \frac{\pi }{d} ((\mu +\eta +\frac{1}{2}) d +n-k)^2}\ 
 \mathrm{e}^{-\kappa  \frac{\pi }{d} ((\mu -\eta +\frac{1}{2}) d +n+k)^2}\\[4mm]
& \, \  +\frac{1}{d}\sum\limits_{k=-s}^s\mathrm{e}^{\frac{4\pi \mathrm{i}}{d}mk}\, 
\sum\limits_{\mu  ,\eta  =-\infty }^\infty \mathrm{e}^{-\kappa  \frac{\pi }{d} ((\mu +\eta +\frac{3}{2}) d +n-k)^2}\ 
 \mathrm{e}^{-\kappa  \frac{\pi }{d} ((\mu -\eta +\frac{1}{2}) d +n+k)^2}\\[4mm]
& \!\!\!\!=\frac{1}{d}\sum\limits_{k=-s}^s\mathrm{e}^{\frac{4\pi \mathrm{i}}{d}mk}\, 
\sum\limits_{\mu  ,\eta  =-\infty }^\infty 
\mathrm{e}^{-2\kappa  \frac{\pi }{d} ((\mu +\frac{1}{2})d +n)^2}\ 
 \mathrm{e}^{-2\kappa  \frac{\pi }{d} (\eta  d -k)^2}\\[4mm]
& \, \ +\frac{1}{d}\sum\limits_{k=-s}^s\mathrm{e}^{\frac{4\pi \mathrm{i}}{d}mk}\, 
\sum\limits_{\mu  ,\eta  =-\infty }^\infty \mathrm{e}^{-2\kappa  \frac{\pi }{d} \left(\mu d +n\right)^2}\ 
 \mathrm{e}^{-2\kappa  \frac{\pi }{d} \left(\left(\eta +\frac{1}{2}\right) d -k\right)^2}\\[4mm]
& \!\!\!\!=\frac{1}{\sqrt{d}}\, \,  \mathfrak{g}_2^{(2\kappa)}(n)\, \,  F[\mathfrak{g}_1^{(2\kappa)}](2m)
 +\frac{1}{\sqrt{d}}\, \, \mathfrak{g}_1^{(2\kappa)}(n)\,\,  F[\mathfrak{g}_2^{(2\kappa)}](2m)\\[4mm]
& \!\!\!\!=\!\frac{1}{\sqrt{2\kappa d}}\, \,  \mathfrak{g}_2^{(2\kappa)}(n) \left( \mathfrak{g}_1^{(2/\kappa)}(m)\!+\!\mathfrak{g}_2^{(2/\kappa)}(m)\right)\!+\!
\frac{1}{\sqrt{2\kappa d}}\,  \mathfrak{g}_1^{(2\kappa)}(n)\!\left( \mathfrak{g}_1^{2/\kappa)} (m)\!-\!\mathfrak{g}_2^{(2/\kappa)}(m)\right).\\[4mm]
\end{array}
\]
c) Since $\mathfrak{g}_3^{(\kappa)}=\frac{1}{\sqrt{\kappa}}\, F[\mathfrak{g}_2^{(1/\kappa)}]$ and $\mathfrak{g}_2^{(1/\kappa)}$ is an even function, from (\ref{FWigner}) we get
\[
W_{\mathfrak{g}_3^{(\kappa)}}(n,m)
=\frac{1}{\kappa}\, W_{F[\mathfrak{g}_2^{(1/\kappa)}]}(n,m)
=\frac{1}{\kappa}\, W_{\mathfrak{g}_2^{(1/\kappa)}}(m,-n). \qquad \opensquare
\]
The Wigner function $W_{\mathfrak{G}_4}$  
is presented in Fig. \ref{bW}, and $W_{\mathfrak{G}_5}(n,m)=W_{\mathfrak{G}_4}(m,-n)$. 

By using the relation (\ref{doublesum}), one obtains
\begin{equation}
\begin{array}{l}
|\mathfrak{g}_1^{(\kappa)}(n)|^2\!=\!\mathfrak{g}_1^{(2\kappa)}(0)\, \mathfrak{g}_1^{(2\kappa)}(n)+\mathfrak{g}_2^{(2\kappa)}(0)\, \mathfrak{g}_2^{(2\kappa)}(n)\\[2mm]
|\mathfrak{g}_2^{(\kappa)}(n)|^2\!=\!\mathfrak{g}_1^{(2\kappa)}(0)\, \mathfrak{g}_2^{(2\kappa)}(n)+\mathfrak{g}_2^{(2\kappa)}(0)\, \mathfrak{g}_1^{(2\kappa)}(n)\\[2mm]
|\mathfrak{g}_3^{(\kappa)}(n)|^2\!=\!\mathfrak{g}_1^{(2\kappa)}(0)\, \mathfrak{g}_1^{(2\kappa)}(n)-\mathfrak{g}_2^{(2\kappa)}(0)\, \mathfrak{g}_2^{(2\kappa)}(n).
\end{array}
\end{equation}
For $d$ large enough, we have $\mathfrak{g}_1^{(2\kappa)}(0)\!\approx \!1$
and $\mathfrak{g}_2^{(2\kappa)}(0)\!\approx \!0$. In this case
\[
|\mathfrak{g}_1^{(\kappa)}(n)|^2\!\approx \!\mathfrak{g}_1^{(2\kappa)}(n), \qquad |\mathfrak{g}_2^{(\kappa)}(n)|^2\!\approx \! \mathfrak{g}_2^{(2\kappa)}(n), \qquad |\mathfrak{g}_3^{(\kappa)}(n)|^2\!\approx \!\mathfrak{g}_1^{(2\kappa)}(n).
\]
If $\psi $ is an even function, then $W_{\psi}(0,0)=\frac{1}{d}\, ||\psi ||^2.$
Particularly, we have
\begin{equation}
\begin{array}{l}
||\mathfrak{g}_1^{(1)}||^2\!=\!\sqrt{\frac{d}{2}}\left( 
\mathfrak{g}_1^{(2)}(0)\right)^2\!+\! 
2\sqrt{\frac{d}{2}}\ \mathfrak{g}_1^{(2)}(0)\,\mathfrak{g}_2^{(2)}(0)\!-\!
\sqrt{\frac{d}{2}}\left(\mathfrak{g}_2^{(2)}(0)\right)^2 \\[2mm]
||\mathfrak{g}_2^{(1)}||^2\!=\!\sqrt{\frac{d}{2}}\left( 
\mathfrak{g}_1^{(2)}(0)\right)^2\!+\! 
\sqrt{\frac{d}{2}}\left(\mathfrak{g}_2^{(2)}(0)\right)^2 
=||\mathfrak{g}_3^{(1)}||^2.
\end{array}
\end{equation}
The binomial coefficients $C_{2j}^m=\frac{(2j)!}{m!\, (2j-m)!}$ satisfy the relation
\begin{equation}
\begin{array}{l}
(C_{2j}^0)^2+(C_{2j}^1)^2+\cdots +(C_{2j}^{2j})^2=C_{4j}^{2j}
\end{array}
\end{equation}
obtained by evaluating the coefficient of $X^{2j}$ from both sides of the equality 
\begin{equation}
(1\!+\!X)^{2j}(X\!+\!1)^{2j}=(1+X)^{4j}.
\end{equation}
Since $F$ is a unitary transformation, this means that 
\begin{equation}
\begin{array}{l}
||\mathfrak{g}_5||^2=||\mathfrak{g}_4||^2=\frac{1}{2^{4j}}\sum\limits_{n=-j}^j\left(\frac{(2j)!}{(j\!-\!n)!\, (j\!+\!n)!} \right)^2=\frac{1}{2^{4j}}\frac{(4j)!}{((2j)!)^2}.
\end{array}
\end{equation}

\section{Coherent state quantization and finite frame quantization}

The standard coherent states are 
\begin{equation}
|q,p\rangle =\mathcal{D}(q,p)|\Psi _0\rangle
\end{equation}
where 
\begin{equation}
\mathcal{D}(q,p)={\rm e}^{-\frac{\rm i}{2}qp}\, {\rm e}^{{\rm i}p\hat{q}}\,{\rm e}^{-{\rm i}q\hat{p}}
\end{equation}
is the displacement operator and $\Psi _0$ is the normalized Gaussian
\begin{equation}
\begin{array}{l}
\Psi_0(q)=\frac{1}{\sqrt[4]{\pi}}\, {\rm e}^{-\frac{1}{2}q^2}.
\end{array}
\end{equation}
The states $|q,p\rangle $ satisfy the relation $\langle q,p|q,p\rangle \!=\!1$ and the resolution of the identity
\begin{equation}
\frac{1}{2\pi }\int _{\mathbb{R}^2}dq\, dp\, |q,p\rangle\langle q,p|=\mathbb{I}.
\end{equation}
 To each function 
$f:\mathbb{R}^2\longrightarrow \mathbb{C}$ with convergent integral we associate the linear operator
\begin{equation}
A_f=\frac{1}{2\pi }\int _{\mathbb{R}^2}dq\, dp\, f(q,p)\,   |q,p\rangle\langle q,p|.
\end{equation} 
This procedure, called {\em coherent state quantization}, allows us to define new operators or to obtain integral representations for known operators \cite{Gazeau}. For example, we have
\begin{equation}\fl
\hat{q}=\frac{1}{2\pi }\int _{\mathbb{R}^2}dq\, dp\, q\,   |q,p\rangle\langle q,p|\qquad \mbox{and}\qquad \hat{p}=\frac{1}{2\pi }\int _{\mathbb{R}^2}dq\, dp\, p\,   |q,p\rangle\langle q,p|.
\end{equation}
By using the finite displacement operators
\begin{equation}
D(\alpha ,\beta) \!=\!{\rm e}^{-\frac{\pi {\rm i}}{d}\alpha \beta}\ {\rm e}^{\frac{2\pi {\rm i}}{d}\beta Q}\ {\rm e}^{-\frac{2\pi {\rm i}}{d}\alpha P}
\!=\!{\rm e}^{\frac{\pi {\rm i}}{d}\alpha \beta}\ {\rm e}^{-\frac{2\pi {\rm i}}{d}\alpha P}\ {\rm e}^{\frac{2\pi {\rm i}}{d}\beta Q}
\end{equation}
and a normalized finite Gaussian  $\mathfrak{G}_i$  we define in $\ell ^2(\mathbb{Z}_d)$ the states
\begin{equation}
|\alpha ,\beta \rangle _{\!i}=D(\alpha ,\beta)|\mathfrak{G}_i\rangle 
\!=\!{\rm e}^{-\frac{\pi {\rm i}}{d}\alpha \beta}\sum\limits_{n=-j}^j
{\rm e}^{\frac{2\pi {\rm i}}{d}\beta n}\ \mathfrak{G}_i(n\!-\!\alpha )\, |j;n\rangle.
\end{equation}
They satisfy the relation ${}_{i}\!\langle \alpha,\beta|\alpha,\beta\rangle _{\!i}\!=\!1$ and the resolution of the identity
\begin{equation}
\frac{1}{d}\sum_{\alpha,\beta=-j}^j |\alpha ,\beta \rangle _{\!i}{}_{i}\!\langle \alpha,\beta|=\mathbb{I}.
\end{equation}
By using the frame $\{|\alpha,\beta\rangle _{\!i}\}_{\alpha,\beta=-j}^j$ we can associate to each function
\[
f\!:\!\{-j,-j\!+\!1,...,j\!-\!1,j\}\!\times \!\{-j,-j\!+\!1,...,j\!-\!1,j\}\!\longrightarrow \!\mathbb{C}
\]
the linear operator $A_f^{(i)}\!:\!\ell^2(\mathbb{Z}_d)\!\longrightarrow \! \ell^2(\mathbb{Z}_d)$,
\begin{equation}
A_f^{(i)}=\frac{1}{d}\sum_{\alpha,\beta=-j}^jf(\alpha,\beta)\,|\alpha ,\beta \rangle _{\!i}{}_{i}\!\langle \alpha,\beta|.
\end{equation}
This procedure, we call {\em finite frame quantization}, allows us to define some new useful linear operators \cite{CD1,CGV,Gazeau}.

\section{Quantum oscillators with finite-dimensional Hilbert space}

The Fourier transform $\Psi \mapsto \mathcal{F}[\Psi ]$, where
\begin{equation}
\mathcal{F}[\Psi ](p)=\frac{1}{\sqrt{2\pi }} \int_{-\infty }^{\infty} 
{\rm e}^{-{\rm i}pq}\Psi (q)\, {\rm d}q,
\end{equation}
satisfies the relation
\[
\mathcal{F}[\hat{p}\Psi ](p)=\frac{-{\rm i}}{\sqrt{2\pi }} \int_{-\infty }^{\infty} 
{\rm e}^{-{\rm i}pq}\Psi '(q)\, {\rm d}q=p\, \mathcal{F}[\Psi ](p)
\]
which can be written as
\begin{equation}
\mathcal{F}\hat{p}=\hat{q}\mathcal{F}\qquad \mbox{or}\qquad \hat{p}=\mathcal{F}^+\hat{q}\mathcal{F}\qquad \mbox{or}\qquad \hat{q}=\mathcal{F}\hat{p}\mathcal{F}^+.
\end{equation}
By using these formulas, the Hamiltonian of the quantum harmonic oscillator
\begin{equation}\label{defH}
H=\frac{1}{2}\hat{p}^2+\frac{1}{2}\hat{q}^2
=-\frac{1}{2}\frac{{\rm d}^2}{{\rm d}q^2}+\frac{1}{2}q^2\, ,
\end{equation}
can be re-written as
\begin{equation}\label{HF}
\begin{array}{l}
H=\frac{1}{2}\mathcal{F}^+\hat{q}^2\mathcal{F}+\frac{1}{2}\hat{q}^2\qquad \mbox{or}\qquad H=\frac{1}{2}\hat{p}^2+\frac{1}{2}\mathcal{F}\hat{p}^2\mathcal{F}^+.
\end{array}
\end{equation} 

%
\subsection{Fourier oscillator}
The self-adjoint operator
\begin{equation}
H_{{}_{\rm Fourier}}:\ell^2 (\mathbb{Z} _d)\longrightarrow \ell^2 (\mathbb{Z} _d),\qquad   H_{{}_{\rm Fourier}}=\frac{1}{2}F^+Q^2F+\frac{1}{2}Q^2,
\end{equation}
where
\begin{equation}
 (Q\psi )(n)=n\, \psi (n)\qquad \mbox{and}\qquad F[\psi ](k)\!=\!\frac{1}{\sqrt{d}}\sum_{n=-s}^s{\rm e}^{-\frac{2\pi {\rm i}}{d}kn}\, \psi (n),
\end{equation}
is the Hamiltonian of a finite-dimensional version of the quantum harmonic oscillator. 
It is Fourier invariant, that is, we have $FH_{{}_{\rm Fourier}}=H_{{}_{\rm Fourier}}F$.
%
\subsection{Harper oscillator}
The finite-difference operator
\begin{equation} 
P^2\!:\!\ell^2 (\mathbb{Z} _d)\!\longrightarrow \!\ell^2 (\mathbb{Z} _d),\qquad  ( P^2\psi )(n)\!=\!-[\psi(n\!+\!1)\!-\!2\psi(n)\!+\!\psi(n\!-\!1)]
\end{equation}
is a finite-dimensional version of the differential operator  $\hat{p}^2=-\frac{{\rm d}^2}{{\rm d}q^2}$. Therefore
\begin{equation}
H_{{}_{\rm Harper}}:\ell^2 (\mathbb{Z} _d)\longrightarrow \ell^2 (\mathbb{Z} _d),\qquad   H_{{}_{\rm Harper}}=\frac{1}{2}P^2+\frac{1}{2}FP^2F^+,
\end{equation} 
can be regarded as the Hamiltonian of a finite oscillator. 
It is Fourier invariant, and one can prove that the eigenspaces corresponding to its eigenvalues are one-dimensional. From the relation
\[ 
H_{{}_{\rm Harper}}\psi=\lambda\, \psi \quad \Longrightarrow \quad H_{{}_{\rm Harper}}(F\psi )=F(H_{{}_{\rm Harper}}\psi)=\lambda(F\psi )
\]
it follows that any eigenfunction of $H_{{}_{\rm Harper}}$ is at the same time an eigenfunction of $F$.
The normalized eigenfunctions $h_0$, $h_1$, ..., $h_{2j}$ of $H_{{}_{\rm Harper}}$, considered in the increasing order of the number of sign alternations, satisfy  the relation \cite{Ba,WK}
\begin{equation}
F\, h_n=(-{\rm i})^n\, h_n
\end{equation} 
and are called {\em Harper functions}. They form an orthonormal basis in $\ell^2 (\mathbb{Z} _d)$, and 
\begin{equation}
F=\sum_{n=0}^{2j}(-{\rm i})^n\, |h_n\rangle \langle h_n|
\end{equation}
is the starting point for the definition \cite{Ba,Ca}
\begin{equation}
F^\alpha =\sum_{n=0}^{2j}(-{\rm i})^{n\alpha }\, |h_n\rangle \langle h_n|
\end{equation}
of the {\em fractional Fourier transform}. If $\alpha $ is in a small enough neighbourhood of $1$, then 
\begin{equation}
H_{{}_{\rm Fourier}}^{(\alpha)}=\frac{1}{2}F^{-\alpha}Q^2F^\alpha +\frac{1}{2}Q^2,\qquad H_{{}_{\rm Harper}}^{(\alpha)}=\frac{1}{2}P^2+\frac{1}{2}F^\alpha P^2F^{-\alpha}
\end{equation}
are deformed versions of the finite oscillators $H_{{}_{\rm Fourier}}$ and $H_{{}_{\rm Harper}}$.

%
\subsection{The Kravchuk oscillator}
Atakishiyev and Wolf proved \cite{AW,HW} that  
\begin{equation}
\begin{array}{l}
H_{{}_ {\rm Kravchuk}}=J_z\!+\!j\!+\!\frac{1}{2}
\end{array}
\end{equation}
can be regarded as  the Hamiltonian of a finite oscillator with coordinate $J_x$ and momentum $-J_y$. 
One can remark the analogy existing between the formulas
\begin{equation}\label{Haa}
\begin{array}{l}
a=\frac{1}{\sqrt{2}}\left(\hat{q}\!+\!{\rm i}\hat{p} \right)\\[1mm]
a^+=\frac{1}{\sqrt{2}}\left(\hat{q}\!-\!{\rm i}\hat{p} \right)\\[1mm]
[H,a]=-a\\[1mm]
[H,a^+]=a^+\\[1mm]
H=a^+a+\frac{1}{2}
\end{array}
\qquad \mbox{and}\qquad 
\begin{array}{l}
J_-=J_x-{\rm i}\, J_y\\[1mm]
J_+=J_x+{\rm i}\, J_y\\[1mm]
[H_{{}_ {\rm Kravchuk}},J_-]=-J_-\\[1mm]
[H_{{}_ {\rm Kravchuk}},J_+]=J_+\\[1mm]
H_{{}_ {\rm Kravchuk}}=\frac{1}{2}(J_+J_-\!-\!J_-J_+)\!+\!j\!+\!\frac{1}{2}.
\end{array}
\end{equation}
From the relation
\begin{equation}
\begin{array}{l}
H_{{}_ {\rm Kravchuk}}|j;n\rangle =\left(n\!+\!j\!+\!\frac{1}{2}\right)|j;n\rangle 
\end{array}
\end{equation}
it follows that the eigenvalues of $H_{{}_ {\rm Kravchuk}}$
are $\frac{1}{2}$,  $1$, $\frac{3}{2}$,  ... , $d\!+\!\frac{1}{2}$ and the corresponding eigenstates are $|j;-j\rangle $, $|j;-j\!+\!1\rangle $, ... , $|j;j\!-\!1\rangle $, $|j;j\rangle $. The relation
\begin{equation}
|j;n\rangle =\sum_{m=-j}^j|\mathfrak{K}_{m}\rangle \langle \mathfrak{K}_{m}|j;n\rangle =\sum_{m=-j}^j\mathfrak{K}_n(m)\, |\mathfrak{K}_{m}\rangle 
\end{equation}
presents the expansion of $|j;n\rangle $ in terms of the coordinate eigenbasis $\{ |\mathfrak{K}_{-m}\rangle \}_{m=-j}^j$.
%

\begin{figure}
\centering
\includegraphics[scale=0.7]{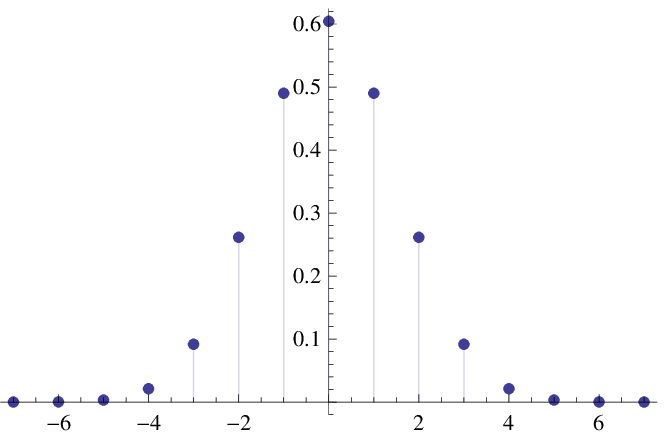}\quad 
\includegraphics[scale=0.7]{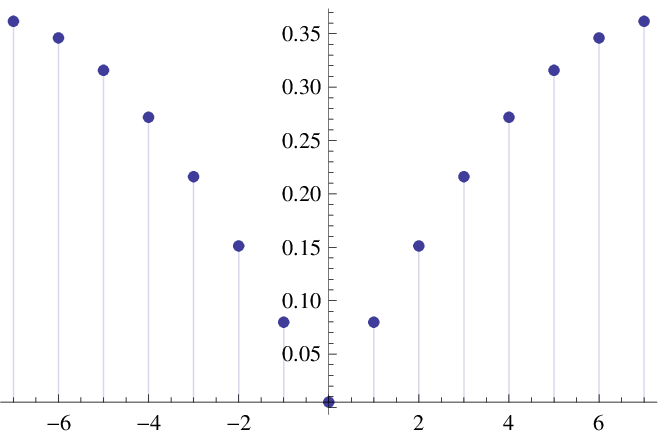}\quad 
\includegraphics[scale=0.7]{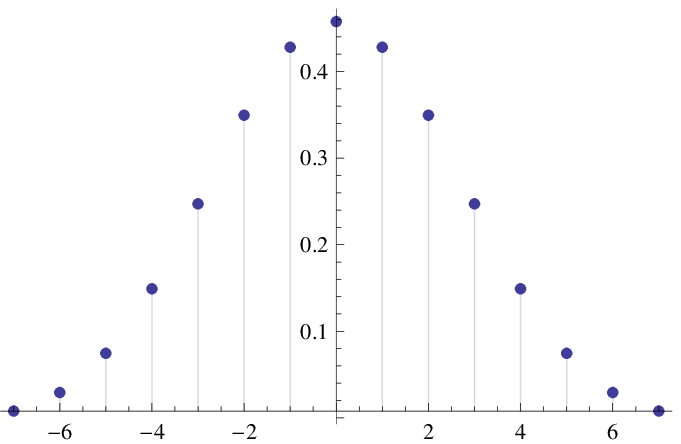}\\[5mm]
\includegraphics[scale=0.75]{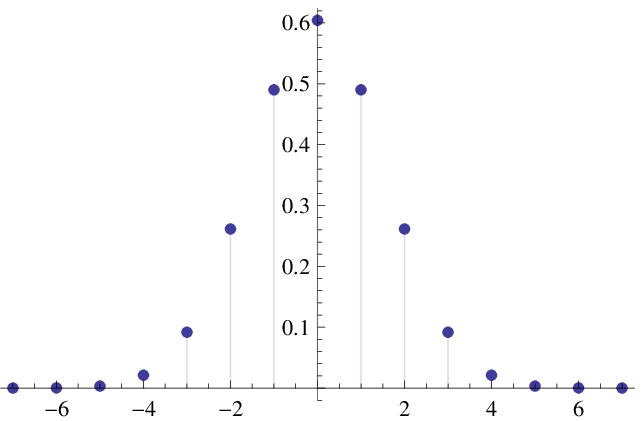}\quad 
\includegraphics[scale=0.7]{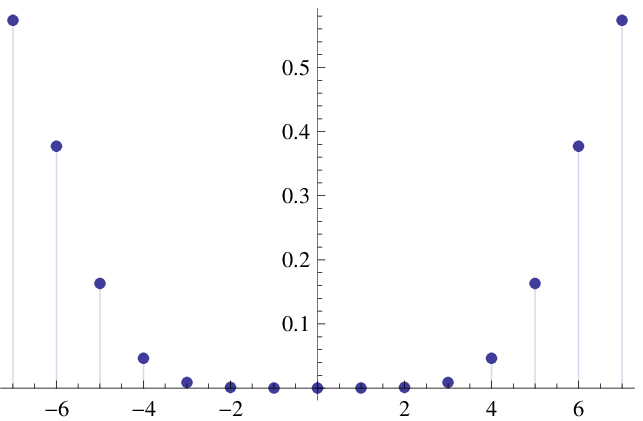}\quad 
\includegraphics[scale=0.75]{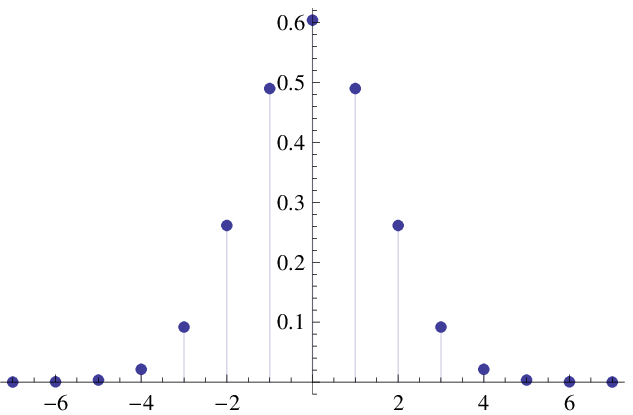}
\caption{\label{ground} The ground state of $H_{{}_{\rm Fourier}}$ (top left), $H_{{}_{\rm Harper}}$ (top center), $H_{{}_{\rm Kravchuk}}$ (top right), ${H}_1$ (bottom left) , ${H}_2$ (bottom center) and ${H}_4$ (bottom right) in the case $d\!=\!15$. }
\end{figure}

%
\subsection{Oscillators defined by using the finite frame quantization}
The Hamiltonian of the harmonic oscillator admits the integral representation \cite{Gazeau}
\begin{equation}\label{HCSQ}
H=-\frac{1}{2}+\frac{1}{2\pi }\int_{\mathbb{R}^2}dq\, dp\left( \frac{q^2}{2}\!+\!\frac{p^2}{2}\right)  |q,p\rangle\langle q,p|.
\end{equation}
Since the frame $\{|\alpha,\beta\rangle _{\!i}\}_{\alpha,\beta=-j}^j$ is a finite counterpart of the standard system of coherent states
$\{ |q,p\rangle \}_{q,p\in \mathbb{R}}$, the linear operator
\begin{equation}
H_i=-\frac{1}{2}+\frac{1}{d}\sum\limits_{\alpha,\beta=-j}^j\left(\frac{\alpha^2}{2}+\frac{\beta^2}{2}\right)|\alpha ,\beta \rangle _{\!i}{}_{i}\!\langle \alpha,\beta|
\end{equation}
is the Hamiltonian of a finite oscillator, for any $i\!\in\!\{1,2,...,5\}$.
From the equality
\begin{equation}
FD(\alpha,\beta)=D(\beta,-\alpha)F
\end{equation}
it follows the relation
\begin{equation}
F|\alpha ,\beta \rangle _{\!i}=FD(\alpha,\beta)|\mathfrak{G}_i\rangle =D(\beta,-\alpha)F|\mathfrak{G}_i\rangle 
\end{equation}
leading to
\[
F|\alpha ,\beta \rangle _{\!1}\!=\!|\beta,-\alpha \rangle _{\!1},\qquad 
F|\alpha ,\beta \rangle _{\!2}\!=\!|\beta,-\alpha \rangle _{\!3},\qquad 
F|\alpha ,\beta \rangle _{\!4}\!=\!|\beta,-\alpha \rangle _{\!5}
\]
and
\begin{equation}
FH_1F^+=H_1,\qquad FH_2F^+=H_3,\qquad FH_4F^+=H_5.
\end{equation}

\begin{center}
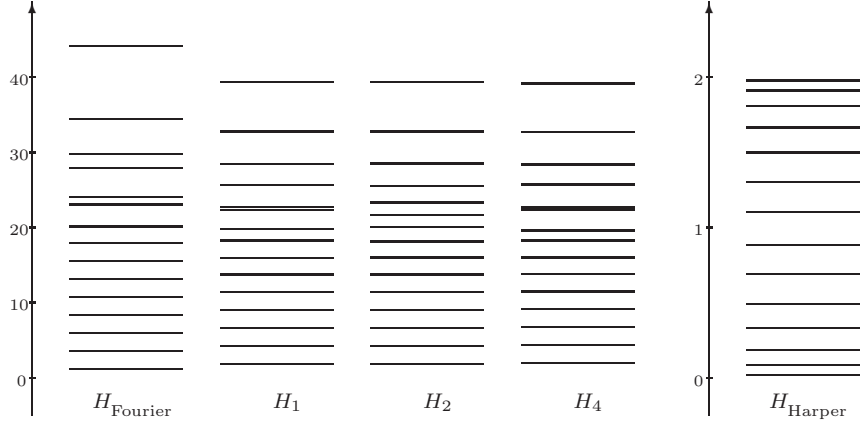
\begin{figure}[h] 
\setlength{\unitlength}{1mm}
\begin{picture}(120,55)(-15,5)

\put(10,5){\vector(0,1){55}}
\put(100,5){\vector(0,1){55}}

\put(9.7,10){\line(1,0){0.7}}
\put(9.7,20){\line(1,0){0.7}}
\put(9.7,30){\line(1,0){0.7}}
\put(9.7,40){\line(1,0){0.7}}
\put(9.7,50){\line(1,0){0.7}}

\put(99.7,10){\line(1,0){0.7}}
\put(99.7,30){\line(1,0){0.7}}
\put(99.7,50){\line(1,0){0.7}}

\put(15,11.19){\line(1,0){15}}
\put(15,13.58){\line(1,0){15}}
\put(15,15.96){\line(1,0){15}}
\put(15,18.35){\line(1,0){15}}
\put(15,20.74){\line(1,0){15}}
\put(15,23.13){\line(1,0){15}}
\put(15,25.50){\line(1,0){15}}
\put(15,27.94){\line(1,0){15}}
\put(15,30.13){\line(1,0){15}}
\put(15,33.06){\line(1,0){15}}
\put(15,34.08){\line(1,0){15}}
\put(15,37.88){\line(1,0){15}}
\put(15,39.80){\line(1,0){15}}
\put(15,44.48){\line(1,0){15}}
\put(15,54.12){\line(1,0){15}}

\put(35,11.88){\line(1,0){15}}
\put(35,14.27){\line(1,0){15}}
\put(35,16.66){\line(1,0){15}}
\put(35,19.04){\line(1,0){15}}
\put(35,21.41){\line(1,0){15}}
\put(35,23.75){\line(1,0){15}}
\put(35,26.00){\line(1,0){15}}
\put(35,28.27){\line(1,0){15}}
\put(35,29.80){\line(1,0){15}}
\put(35,32.35){\line(1,0){15}}
\put(35,32.73){\line(1,0){15}}
\put(35,35.66){\line(1,0){15}}
\put(35,38.46){\line(1,0){15}}
\put(35,42.80){\line(1,0){15}}
\put(35,49.35){\line(1,0){15}}

\put(55,11.88){\line(1,0){15}}
\put(55,14.27){\line(1,0){15}}
\put(55,16.66){\line(1,0){15}}
\put(55,19.04){\line(1,0){15}}
\put(55,21.41){\line(1,0){15}}
\put(55,23.75){\line(1,0){15}}
\put(55,26.02){\line(1,0){15}}
\put(55,28.16){\line(1,0){15}}
\put(55,30.07){\line(1,0){15}}
\put(55,31.69){\line(1,0){15}}
\put(55,33.34){\line(1,0){15}}
\put(55,35.49){\line(1,0){15}}
\put(55,38.50){\line(1,0){15}}
\put(55,42.80){\line(1,0){15}}
\put(55,49.35){\line(1,0){15}}

\put(75,11.99){\line(1,0){15}}
\put(75,14.38){\line(1,0){15}}
\put(75,16.76){\line(1,0){15}}
\put(75,19.14){\line(1,0){15}}
\put(75,21.49){\line(1,0){15}}
\put(75,23.81){\line(1,0){15}}
\put(75,26.01){\line(1,0){15}}
\put(75,28.25){\line(1,0){15}}
\put(75,29.62){\line(1,0){15}}
\put(75,32,29){\line(1,0){15}}
\put(75,32.67){\line(1,0){15}}
\put(75,35.71){\line(1,0){15}}
\put(75,38.40){\line(1,0){15}}
\put(75,42.73){\line(1,0){15}}
\put(75,49.18){\line(1,0){15}}

\put(105,10.42){\line(1,0){15}}
\put(105,11.72){\line(1,0){15}}
\put(105,13.75){\line(1,0){15}}
\put(105,16.61){\line(1,0){15}}
\put(105,19.86){\line(1,0){15}}
\put(105,23.81){\line(1,0){15}}
\put(105,27.73){\line(1,0){15}}
\put(105,32.09){\line(1,0){15}}
\put(105,36.01){\line(1,0){15}}
\put(105,40.0){\line(1,0){15}}
\put(105,43.28){\line(1,0){15}}
\put(105,46.18){\line(1,0){15}}
\put(105,48.24){\line(1,0){15}}
\put(105,49.56){\line(1,0){15}}

\put(8,9){$\scriptscriptstyle{0}$}
\put(7,19){$\scriptscriptstyle{10}$}
\put(7,29){$\scriptscriptstyle{20}$}
\put(7,39){$\scriptscriptstyle{30}$}
\put(7,49){$\scriptscriptstyle{40}$}

\put(98,9){$\scriptscriptstyle{0}$}
\put(98,29){$\scriptscriptstyle{1}$}
\put(98,49){$\scriptscriptstyle{2}$}

\put(18,6){$\scriptstyle{H_{{}_{\rm Fourier}}}$}
\put(42,6){$\scriptstyle{{H}_1}$}
\put(62,6){$\scriptstyle{{H}_2}$}
\put(82,6){$\scriptstyle{{H}_4}$}
\put(108,6){$\scriptstyle{H_{{}_{\rm Harper}}}$}

\end{picture}
\caption{\label{eigval} The eigenvalues of $H_{{}_{\rm Fourier}}$, ${H}_1$, ${H}_2$, ${H}_4$  and $H_{{}_{\rm Harper}}$  in the case $d\!=\!15$.}
\end{figure}
\end{center}

\subsection{Oscillators defined by using the Gramm-Schmidt method of orthonormalization }
The normalized Kravchuk polynomials  $\Phi_m(X)=\sqrt{\frac{1}{C_{2j}^{j+m}}}\ K_m(X)$ can be obtained by orthonormalizing 
$1,\, X,\, ...,\, X^{2j}$ with respect to the scalar product
\begin{equation}
\langle \varphi ,\psi \rangle =\sum\limits_{n=-j}^j \mathfrak{g}_4(n)\, \varphi(n)\, \psi(n),
\end{equation}
and the Kravchuk functions can be defined as
\begin{equation}
\mathfrak{K}_m(n)=\sqrt{\mathfrak{g}_4(n)} \ \Phi_m(n).
\end{equation}

Let $i\!\in\!\{ 1,2,3,4,5\}$ and let $\Phi_{-j}^{(i)}$, $\Phi_{-j+1}^{(i)}$, ..., $\Phi_{j}^{(i)}$ be the polynomials obtained by orthonormalizing 
$1,\, X,\, ...,\, X^{2j}$ with respect to the scalar product
\begin{equation}
\langle \varphi ,\psi \rangle =\sum\limits_{n=-j}^j \mathfrak{G}_i^2(n)\, \varphi(n)\, \psi(n)
\end{equation}
by using the Gramm-Schmidt method of orthonormalization.  The functions 
\begin{equation}
\phi _m^{(i)}(n)=\mathfrak{G}_i(n) \ \Phi_m^{(i)}(n)
\end{equation}
represent a finite counterpart of the Hermite-Gauss functions $\Psi _n$, and
\begin{equation}
\begin{array}{l}
\tilde H_i=\sum\limits_{m=-j}^j(j\!+\!m\!+\!\frac{1}{2})
|\phi _m^{(i)}\rangle \langle \phi _m^{(i)}|
\end{array}
\end{equation}
a finite counterpart of the relation
\begin{equation}\label{HHG}
\begin{array}{l}
H=\sum\limits_{n=0}^\infty \left(n\!+\!\frac{1}{2}\right) |\Psi_n\rangle\langle\Psi_n|
\end{array}
\end{equation}
satisfied by the Hamiltonian of the harmonic oscillator. Therefore, the operator $\tilde H_i$ can be regarded as the Hamiltonian of a finite oscillator with the ground state $\mathfrak{G}_i$.

\begin{table}
\caption{ \label{tabFK} The eigenvectors of $F$ and the Kravchuk functions in the case $d\!=\!3$.}
\begin{indented}
\lineup
\item[]\begin{tabular}{rrrrrrr}
\br
 $n$ \ & $\mathfrak{F}_{-1}(n)$ \ \ & \ $\mathfrak{F}_0(n)$ & \ $\mathfrak{F}_1(n)$ \ \ \  & \ $\mathfrak{K}_{-1}(n)$ & \ $\mathfrak{K}_0(n)$ & \ $\mathfrak{K}_1(n)$  \\
\mr
$-1$  & $\frac{1}{2}\sqrt{1\!-\!\frac{1}{\sqrt{3}}}$ & $-\frac{1}{\sqrt{2}}$ & $\frac{1}{2}\sqrt{1\!+\!\frac{1}{\sqrt{3}}}$ & $\frac{1}{2}$ \ \ \ & $\frac{1}{\sqrt{2}}$ & $\frac{1}{2}$ \ \\
$0$ \ & \ $\frac{1}{\sqrt{2}}\sqrt{1\!+\!\frac{1}{\sqrt{3}}}$ & $0$ \  & $-\frac{1}{\sqrt{2}}\sqrt{1\!-\!\frac{1}{\sqrt{3}}}$  & $\frac{1}{\sqrt{2}}$  \ \ & $0$ \  & $-\frac{1}{\sqrt{2}}$ \\
$1$ \ & $\frac{1}{2}\sqrt{1\!-\!\frac{1}{\sqrt{3}}}$ & $\frac{1}{\sqrt{2}}$ & 
$\frac{1}{2}\sqrt{1\!+\!\frac{1}{\sqrt{3}}}$  & $\frac{1}{2}$\ \ \ \ & $-\frac{1}{\sqrt{2}}$ & 
$\frac{1}{2}$ \ \\
\br
\end{tabular}
\end{indented}
\end{table}

\section{On the existence of revivals}

When the dimension $d$ becomes larger and larger, the eigenvalues of the considered finite oscillators have the tendency to become equidistant (see Fig. \ref{eigval}). On the other hand, the existence of some equidistant energy levels implies the existence of revivals.
If the Hamiltonian $H_{{}_{\rm finite}}$ of a finite oscillator admits $k\geq 3$ eigenvalues $\varepsilon _1$, $\varepsilon _2$, ..., $\varepsilon _m$ such that
\[
\varepsilon _2-\varepsilon _1=\varepsilon _3-\varepsilon _2=...=\varepsilon _k-\varepsilon _{k-1}
\]
then, by denoting with $|\psi _1\rangle $, $|\psi _2\rangle $, ..., $|\psi _k\rangle $
the corresponding eigenvectors, we have 
\begin{equation}\fl
\begin{array}{rl}
{\rm e}^{-{\rm i}tH_{{}_{\rm finite}}}|\psi \rangle & \!\!\!=\alpha _1{\rm e}^{-{\rm i}t\varepsilon _1} |\psi _1\rangle +\alpha _2 {\rm e}^{-{\rm i}t\varepsilon _2} |\psi _2\rangle +...+\alpha _k {\rm e}^{-{\rm i}t\varepsilon _k} |\psi _k\rangle  \\[2mm]
 & \!\!\!={\rm e}^{-{\rm i}t\varepsilon _1}\left(\alpha _1 |\psi _1\rangle +\alpha _2 {\rm e}^{-{\rm i}t(\varepsilon _2-\varepsilon _1)} |\psi _2\rangle +...+\alpha _k {\rm e}^{-{\rm i}t(\varepsilon _k-\varepsilon _1)} |\psi _k\rangle  \right)\\[2mm]
 & \!\!\!={\rm e}^{-{\rm i}t\varepsilon _1}\left(\alpha _1 |\psi _1\rangle \!+\!\alpha _2 {\rm e}^{-{\rm i}t(\varepsilon _2-\varepsilon _1)} |\psi _2\rangle \!+...+\!\alpha _k {\rm e}^{-{\rm i}t(k-1)(\varepsilon _2-\varepsilon _1)} |\psi _k\rangle  \right)
\end{array}
\end{equation}
for any state $|\psi \rangle $ of the form
\[
|\psi \rangle =\alpha _1 \, |\psi _1\rangle +\alpha _2 \, |\psi _2\rangle + ...+\alpha _k \, |\psi _k\rangle.
\]
Particularly, up to a phase factor, we have the equality 
\begin{equation}
{\rm e}^{-{\rm i}\left(t+\frac{2\pi }{\varepsilon _2-\varepsilon _1}\right)H_{{}_{\rm finite}}}|\psi \rangle = {\rm e}^{-{\rm i}tH_{{}_{\rm finite}}}|\psi \rangle 
\end{equation}
which shows that the time evolution is periodic, and the number $\frac{2\pi }{\varepsilon _2-\varepsilon _1}$ is a period for any coefficients $\alpha _1$, $\alpha _2$, ... , $\alpha _k$.

\section{Gaussians and oscillators in a three-dimensional Hilbert space}

The Fourier transform has in the computational basis $\{ |1;-1\rangle,\, |1;0\rangle,\, |1;1\rangle \}$ the matrix 
\begin{equation}
F=\frac{1}{2\sqrt{3}}\left(
\begin{array}{ccc}
-1\!-\!\sqrt{3}\, {\rm i} \ & 2 \ & -1\!+\!\sqrt{3}\, {\rm i}\\
2 & 2 & 2\\
-1\!+\!\sqrt{3}\, {\rm i} & 2 & -1\!-\!\sqrt{3}\, {\rm i}
\end{array}\right)
\end{equation}
and its eigenvalues are 1, $-{\rm i}$, -1. \ It admits the spectral decomposition
\begin{equation}
F=|\mathfrak{F}_{-1}\rangle \langle \mathfrak{F}_{-1}|-{\rm i}\, |\mathfrak{F}_{0}\rangle \langle \mathfrak{F}_{0}|-|\mathfrak{F}_{1}\rangle \langle \mathfrak{F}_{1}|,
\end{equation}
where $|\mathfrak{F}_{-1}\rangle$, $|\mathfrak{F}_{0}\rangle$, $|\mathfrak{F}_{1}\rangle$ are the normalized eigenvectors presented in Table \ref{tabFK}. \\
The finite Gaussians $\mathfrak{G}_1(n)$, $\mathfrak{G}_2(n)$, $\mathfrak{G}_3(n)$, obtained by using the relations
\begin{equation}\fl
\begin{array}{l}
F[\mathfrak{G}_1]=\mathfrak{G}_1,\qquad F[\mathfrak{G}_2\!+\!\mathfrak{G}_3]=\mathfrak{G}_2\!+\!\mathfrak{G}_3,\qquad F[\mathfrak{G}_2\!-\!\mathfrak{G}_3]=-(\mathfrak{G}_2\!-\!\mathfrak{G}_3)
\end{array}
\end{equation}
are  presented in Table \ref{tabG}.
The Kravchuk transform has the matrix
\begin{equation}
K=\frac{1}{2}\left(
\begin{array}{ccc}
1 \ & \sqrt{2} \ & 1\\
-\sqrt{2} & 0 & \sqrt{2}\\
1 & -\sqrt{2} & 1
\end{array}\right),
\end{equation}
and we have $K|1;-1\rangle \!=\!|\mathfrak{K}_1\rangle $, \ $K|1;0\rangle \!=\!|\mathfrak{K}_0\rangle $, \ $K|1;1\rangle \!=\!|\mathfrak{K}_{-1}\rangle $, 
\begin{equation}
\begin{array}{l}
J_z=\left(
\begin{array}{ccc}
-1 \ & 0 \ & 0\\
0 & 0 & 0\\
0 & 0 & 1
\end{array}\right)=-|1;-1\rangle \langle 1;-1|+|1;1\rangle \langle 1;1|\\[8mm]
J_x=\frac{1}{2}\left(
\begin{array}{ccc}
0 \ & \sqrt{2} \ & 0\\
\sqrt{2} & 0 & \sqrt{2}\\
0 & \sqrt{2} & 0
\end{array}\right)=KJ_zK^+=-|\mathfrak{K}_1\rangle \langle \mathfrak{K}_1|+|\mathfrak{K}_{-1}\rangle \langle \mathfrak{K}_{-1}|.
\end{array}
\end{equation}
The eigenvectors of the Fourier transform are at the same time eigenvectors of the Fourier invariant Hamiltonians $H_{{}_{\rm Fourier}}$, $H_{{}_{\rm Harper}}$ and $H_1$. More exactly, we have
\begin{equation}
\begin{array}{r}
H_{{}_{\rm Fourier}}=\frac{1}{2}\left(1\!-\!\frac{1}{\sqrt{3}}\right) |\mathfrak{F}_{-1}\rangle \langle \mathfrak{F}_{-1}|\, +\, |\mathfrak{F}_{0}\rangle \langle \mathfrak{F}_{0}|+\frac{1}{2}\left(1\!+\!\frac{1}{\sqrt{3}}\right)  |\mathfrak{F}_{1}\rangle \langle \mathfrak{F}_{1}|\\[3mm]
H_{{}_{\rm Harper}}=\frac{1}{2}\left(1\!-\!\frac{1}{\sqrt{3}}\right) |\mathfrak{F}_{-1}\rangle \langle \mathfrak{F}_{-1}|\!+\!3\, |\mathfrak{F}_{0}\rangle \langle \mathfrak{F}_{0}|+\frac{1}{2}\left(1\!+\!\frac{1}{\sqrt{3}}\right)  |\mathfrak{F}_{1}\rangle \langle \mathfrak{F}_{1}|\\[3mm]
H_1=\frac{1}{2}\left(1\!-\!\frac{1}{2\sqrt{3}}\right)\ |\mathfrak{F}_{-1}\rangle \langle \mathfrak{F}_{-1}|\!+\!\frac{1}{4}\left(3\!+\!\frac{1}{\sqrt{3}}\right)\, |\mathfrak{F}_{0}\rangle \langle \mathfrak{F}_{0}|+\frac{3}{4}\ |\mathfrak{F}_{1}\rangle \langle \mathfrak{F}_{1}|.
\end{array}
\end{equation}
These operators and the functions $\mathfrak{F}_{i}$, $\mathfrak{K}_{i}$, $\mathfrak{G}_{i}$   may be useful in the study of qutrits.

\begin{table}
\caption{ \label{tabG}The finite Gaussians $\mathfrak{G}_i$ in the case $d\!=\!3$.}
\begin{indented}
\lineup
\item[]\begin{tabular}{rrrrrr}
\br
 $n$ \ & $\mathfrak{G}_1(n)$\ \ \ \  & $\mathfrak{G}_2(n)$\ \ \ \  & $\mathfrak{G}_3(n)$ \ \ \ \  & $\mathfrak{G}_4(n)$ & $\mathfrak{G}_5(n)$ \\
\mr
$-1$  & $\frac{1}{2}\sqrt{1\!-\!\frac{1}{\sqrt{3}}}$ & $\frac{1}{2}\sqrt{1\!+\!\sqrt{\frac{2}{3}}}$ & $-\frac{1}{2}\sqrt{1\!-\!\sqrt{\frac{2}{3}}}$  & $\frac{1}{\sqrt{6}}$ \ \  & $\frac{1}{3\sqrt{2}}$ \ \\
$0$ \ & $\frac{1}{\sqrt{2}}\sqrt{1\!+\!\frac{1}{\sqrt{3}}}$ & $\frac{1}{\sqrt{2}}\sqrt{1\!-\!\sqrt{\frac{2}{3}}}$  & $\frac{1}{\sqrt{2}}\sqrt{1\!+\!\sqrt{\frac{2}{3}}}$  & $\frac{2}{\sqrt{6}}$ \ \  & $\frac{4}{3\sqrt{2}}$ \ \\
$1$ \ & $\frac{1}{2}\sqrt{1\!-\!\frac{1}{\sqrt{3}}}$ & $\frac{1}{2}\sqrt{1\!+\!\sqrt{\frac{2}{3}}}$ & 
$-\frac{1}{2}\sqrt{1\!-\!\sqrt{\frac{2}{3}}}$ & $\frac{1}{\sqrt{6}}$ \ \  & $\frac{1}{3\sqrt{2}}$ \ \\
\br
\end{tabular}
\end{indented}
\end{table}

\section{Concluding remarks}

There exists a close relation between the finite Gaussians and finite oscillators.
We have used certain finite Gaussians in order to define some finite oscillators.
Conversely, the ground state of a finite oscillator (see Fig. \ref{ground}) can be regarded as a finite version of the  Gaussian function. In the case of the Fourier invariant Hamiltonians $H_{{}_{\rm Fourier}}$, $H_{{}_{\rm Harper}}$ and $H_1$,
the ground state is an eigenfunction of $F$.
The  normalized eigenfunction
\begin{equation}
\begin{array}{l}
\frac{1}{\sqrt{2}}\left(\mathfrak{G}_4\!+\!\mathfrak{G}_5 \right)(n)=
\frac{2^{2j}}{\sqrt{2}}\frac{(2j)!}{\sqrt{(4j)!}}\left(\frac{1}{2^{2j}}\frac{(2j)!}{(j\!-\!n)!\, (j\!+\!n)!}+
\frac{1}{\sqrt{d}}\, \cos^{2j}\frac{n\pi }{d}\right)
\end{array}
\end{equation}
of $F$ can also be regarded as a finite version of the Gaussian function $\Psi_0(q)=\frac{1}{\sqrt[4]{\pi}}\, {\rm e}^{-\frac{1}{2}q^2}$.

\section*{References}

\end{document}